\documentclass[fleqn,usenatbib]{mnras}

\usepackage[T1]{fontenc}
\usepackage{ae,aecompl}
 
\usepackage{graphicx}	% Including figure files
\usepackage{amsmath}	% Advanced maths commands

\usepackage{amssymb}	% Extra maths symbols
\usepackage{float}
\usepackage{caption}
\usepackage{subcaption}

\title[Radial profiles of the Centaurus Cluster]{ Chemical enrichment of ICM within the Centaurus cluster I: radial profiles.}

\author[Gatuzz et al.]{
Efrain Gatuzz$^{1}$\thanks{E-mail: egatuzz@mpe.mpg.de},
J. S. Sanders$^{1}$,
K. Dennerl$^{1}$,
A. Liu$^{1}$,
A. C. Fabian$^{2}$,
C. Pinto$^{3}$,\newauthor
D. Eckert$^{4}$,
S. A. Walker$^{5}$
and J. ZuHone$^{6}$
\\
$^{1}$ Max-Planck-Institut f\"ur extraterrestrische Physik, Gie{\ss}enbachstra{\ss}e 1, 85748 Garching, Germany\\
$^{2}$ Institute of Astronomy, Madingley Road, Cambridge CB3 0HA, UK\\ 
$^{3}$ INAF - IASF Palermo, Via U. La Malfa 153, I-90146 Palermo, Italy \\
$^{4}$ Department of Astronomy, University of Geneva, Ch. d\rq Ecogia 16, CH-1290 Versoix, Switzerland \\ 
$^{5}$ Department of Physics and Astronomy, University of Alabama in Huntsville, Huntsville, AL 35899, USA\\
$^{6}$ Harvard-Smithsonian Center for Astrophysics, 60 Garden Street, Cambridge, MA, 02138, USA
}

\date{Accepted XXX. Received YYY; in original form ZZZ} 
\pubyear{2023} 
\begin{document}
 \label{firstpage}
\pagerange{\pageref{firstpage}--\pageref{lastpage}}
\maketitle 

\begin{abstract}
We examine deep {\it XMM-Newton} EPIC-pn observations of the Centaurus cluster to study the hot intracluster medium (ICM) and radial metal distributions within such an environment. 
We found that the best-fit spectral model corresponds to a log-normal temperature distribution, with discontinuities around $\sim10$~kpc, $\sim50$~kpc, and $\sim100$~kpc, also observed in the abundances distributions. 
We measured the radial profiles of O, Si, S, Ar, Ca, and Fe. These profiles reveal prominent negative gradients for distances $<90$~kpc, which then transition to flatter profiles.
We modeled X/Fe ratio profiles with a linear combination of SNIcc and SNIa models. 
The best-fit model suggests a uniform SNIa percentage contribution to the total cluster enrichment, thus supporting an early enrichment of the ICM, with most of the metals present being produced before clustering.

\end{abstract}


\begin{keywords}
X-rays: galaxies: clusters -- galaxies: clusters: general -- galaxies: clusters: intracluster medium -- galaxies: clusters: individual: Centaurus
\end{keywords}

\section{Introduction}\label{sec_in}   
The distribution of metals in the diffuse intracluster medium (ICM) provides invaluable information on the origin of heavy elements, which are synthesized by supernovae in galaxy clusters, as well as the formation and evolutionary history of galaxies. 
Light $\alpha$-elements (O, Ne, Mg) are mainly produced in core-collapse supernovae (SNcc), while Fe-peak elements (Cr, Mn, Fe, Ni) mainly originate from type~Ia supernovae (SNIa). 
Intermediate-mass elements (e.g., Si, S, Ar, and Ca) are synthesized by both SNcc and SNIa \citep[e.g.,][and references therein]{nom13}. 
Other essential parameters that regulate the metal production include the initial metallicity of the progenitors, the initial mass function (IMF) of the stars that explode as SNcc, and the SNIa explosion mechanism \citep[see][for a review]{wer08}. 
Regarding the explosion mechanisms, the two main theoretical models proposed include delayed denotation and pure deflagration.
In the first case, a subsonic burning process starts at the center of the white dwarf due to the ignition of nuclear reactions. 
As the deflagration propagates outward, it encounters a region of lower density where a transition between deflagration and detonation (i.e., a supersonic combustion process) occurs due to shock compression, the accumulation of burned material, and the development of hydrodynamic instabilities. 
This process rapidly releases energy, creating a powerful shock wave that disrupts the progenitor, resulting in an SNIa \citep{hoe96,iwa99,gam05,rop12}.
In the second case, the final explosion is driven entirely by the subsonic combustion process. 
Unlike the delayed detonation mechanism, there is no transition regime in a pure deflagration model.
Instead, the deflagration front continues to propagate outward, consuming more and more material and releasing additional energy causing the white dwarf to expand rapidly and disrupt itself \citep{ple04,ple07,kas07,fin14,lon14}.
These competing models are still under debate in the community. However, measuring accurate abundances in the ICM can help to distinguish between both scenarios.

X-ray spectroscopy of the ICM provides a powerful tool to constrain heavy-elements abundances from the intensity of their emission lines  \citep[see][for a recent review]{mer18}. 
Many cool-core clusters have shown a central Fe abundance excess \citep[e.g.][]{deg01,chu03,pan15,mer17,liu19,liu20}. 
Outside the core regions, flat and azimuthally uniform Fe distributions towards the outskirts have been observed \citep{mat11,wer13,sim15,urb17}. 
{\it Hitomi} observations of the Perseus cluster indicated that near the cluster core, the abundance ratios are fully consistent with solar \citep{hit18,sim19}. 
This result points out the importance of the contribution from both near SNcc and SNIa (with sub-Chandrasekhar mass) to the chemical enrichment of the ICM. 
Although simple combinations of SNcc and Ia struggle to describe the abundances. 
Including neutrino physics in the core-collapse supernova, yield calculations may improve the agreement with the observed pattern of $\alpha$-elements in the Perseus Cluster core \citep[see e.g.][]{sim19}.

 The Centaurus cluster, located at $z=0.0104$ \citep{luc86a}, is an excellent target to study the chemical enrichment of ICM. 
Two subgroups are identified using optical observations \citep{luc86a}. 
The main one centered on the galaxy NGC~4696 (refereed as Cen~30), and a second one centered on the galaxy NGC~4709 (refereed as Cen~45) 5 arcmin to the east with a line-of-sight. 
\citet{wal13a} proposed a simple shock-heating model between the two systems. 
Multiple sharp discontinuities in the X-ray surface brightness have been found in this system \citep{san16}. 
Such structures, characterized by gradients in temperature and density across the edges, are known as cold fronts that separate regions of different thermodynamic states within the ICM \citep{mar00,vik01,mar07}. 
They arise as a gas sloshing feature of the ICM during cluster mergers when a dense, cooler subcluster or a cool core encounters the hotter, less dense gas of the main cluster. 
\citet{san16} also found a series of notches around the western cold front, suggesting the presence of Kelvin–Helmholtz instabilities (i.e., due to the interaction between the velocity difference across the interface and the shear forces acting on it) of length scales of $\sim7$~kpc along the edge. 
Such instabilities at cold fronts play a crucial role in the mixing and transport processes within the ICM \citep{roed13}.
 Moreover, \citet{san16} shows that the central AGN appears to have been repeatedly active over long timescales with periods of 10s of Myr. 
A super-solar ($\sim 1.5-2$ solar) value for Fe and intermediate-mass elements has been reported \citep{mat07,tak09,sak11,san16,fuk22}. 
Therefore, it is an ideal target to study the contribution of both SNcc and SNIa to the metal abundances.

We present an analysis of the chemical enrichment of the ICM within the Centaurus galaxy cluster. Four observations with large exposure time ( $>100$~ks) have been used for the first time to analyze the metallicity distribution within this source, thus allowing for better statistics. The outline of this paper is as follows. In Section~\ref{sec_dat}, we describe the data reduction process.  In Section \ref{sec_fits}, we explain the fitting procedure. Section~\ref{sec_dis} shows the results discussed in Section~\ref{sec_dis2}. Finally, the conclusions and summary are included in Section~\ref{sec_con}. Throughout this paper we assumed a $\Lambda$CDM cosmology with $\Omega_m = 0.3$, $\Omega_\Lambda = 0.7$, and $H_{0} = 70 \textrm{ km s}^{-1}\ \textrm{Mpc}^{-1} $.

\section{Data reduction}\label{sec_dat}
We used the same observations analyzed in \citet{gat22b} and followed the reduction process shown in \citet[][c]{san20,gat22a}. 
Four observations with large exposure time ( $>100$~ks) have been used for the first time to analyze the metallicity distribution within this source (ObsID: 0823580701, 0823580401, 0823580801, 0823580301).
 The {\it XMM-Newton} European Photon Imaging Camera \citep[EPIC,][]{str01} spectra were reduced with the Science Analysis System (SAS\footnote{\url{https://www.cosmos.esa.int/web/xmm-newton/sas}}, version 19.1.0). 
We processed each observation with the {\tt epchain} SAS tool. 
Bad time intervals were filtered from flares applying a 1.0 cts/s rate threshold, while we used only single-pixel events (PATTERN==0). 
We filtered the data using FLAG==0 to avoid bad pixels or regions close to CCD edges. 

We created the event files using a new energy calibration scale developed in \citet{san20} to measure line-of-sight velocities with uncertainties down to 100~km/s at the Fe-K complex by using the background X-ray lines identified in the spectra of the detector as references for the absolute energy scale. With such procedure, we have quantified bulk velocities for the Virgo cluster \citet{gat22a}, Centaurus cluster \citet{gat22b}, and Ophiuchus cluster \citet{gat23c}. In this context, we used only EPIC-pn corrected event files to study the emission lines and measure the chemical composition of the ICM. We identified point sources using the SAS task {\tt edetect\_chain}, with a likelihood parameter {\tt det\_ml} $> 10$. Such point sources were excluded from the subsequent analysis. 

We analyzed non-overlapping circular regions to study the distribution of chemical elements in the ICM. 
The thickness of these rings increases as the square root distance from the Centaurus center. Figure~\ref{fig_regions} shows the extracted regions.    
    
\section{Spectral fitting}\label{sec_fits}  
For each spatial region, we combined the spectra from different observations. 
Then, we load the data twice to fit separately but simultaneously the hard (4.0-10 keV) and soft (0.5-4.0 keV)  energy bands. 
The new EPIC-pn energy calibration scale cannot be applied for lower energies \citep[see ][]{san20}. Therefore, we only set the redshift as a free parameter for the 4.0-10 keV energy band. Including lower energy band data leads to a better constraint for the temperatures and metallicities. 
The previous work in \citet{gat22b} did not include $<4$~keV data. 
We analyze the spectra with the {\it xspec} spectral fitting package (version 12.11.1\footnote{\url{https://heasarc.gsfc.nasa.gov/xanadu/xspec/}}). 
We assumed {\tt cash} statistics \citep{cas79}. Errors are quoted at 1$\sigma$ confidence level unless otherwise stated. Abundances are given relative to \citet{lod09}.

We tested different models to fit the spectra, namely (a) a single {\tt apec} thermal component model; (b) a two {\tt apec} thermal components model, and (b) a {\tt lognorm} model. Figure~\ref{cstat_comparison} compares the best-fit statistic obtained for each model. 
The numbering is from the innermost to the outermost. 
We have excluded the central part of the cluster (i.e., the AGN). For all regions but one, the {\tt lognorm} model provides the best-fit statistic while the single-temperature component performs worse. 
Therefore we decided to analyze the Centaurus cluster using the {\tt lognorm} model best-fit results. 
We also note that previous analysis suggests that a lognormal distribution has a more physical meaning when fitting X-ray galaxy clusters \citep{vij22}.   

The {\tt lognorm} model assumes a log-normal temperature distribution and takes as input the width of the temperature distribution in log space ($\sigma$), a central temperature ($kT$), metallicities, redshift, and normalization. 
We included a {\tt tbabs} component \citep{wil00} to account for the Galactic absorption. 
The free parameters of the model are the column density ($N({\rm H})$), temperature, log($\sigma$), elemental abundances (O, Si, S, Ar, Ca, Fe ), and normalization. 
It is important to note that Al and Mg abundances might not be reliable due to a strong Al K$\alpha$ instrumental line $\sim 1.5$ keV. 
Similarly, degeneracies with the log($\sigma$) parameter and the multi-temperature modeling could affect the Ni and Ne abundances. 
Moreover, a first attempt at fitting the data with Ni and Ne abundances as free parameters led to an irregular distribution of them and a significant scattering of log($\sigma$). 
Therefore, we decided to fix these abundances to the Fe value.

\begin{figure}    
\centering
\includegraphics[width=0.49\textwidth]{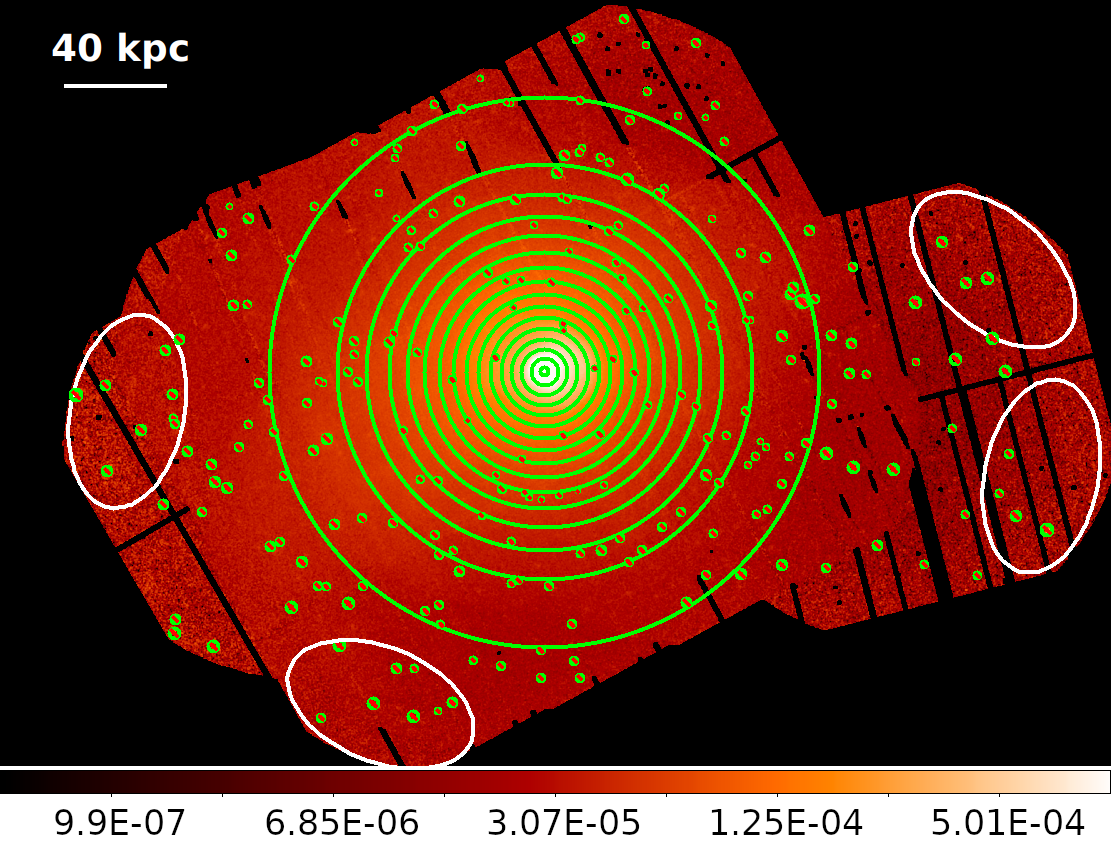} 
\caption{Centaurus cluster extracted regions. Black circles correspond to point sources which were excluded from the  analysis. The white regions were used to model the astrophysical background.} \label{fig_regions} 
\end{figure}

\begin{figure}    
\centering
\includegraphics[width=0.49\textwidth]{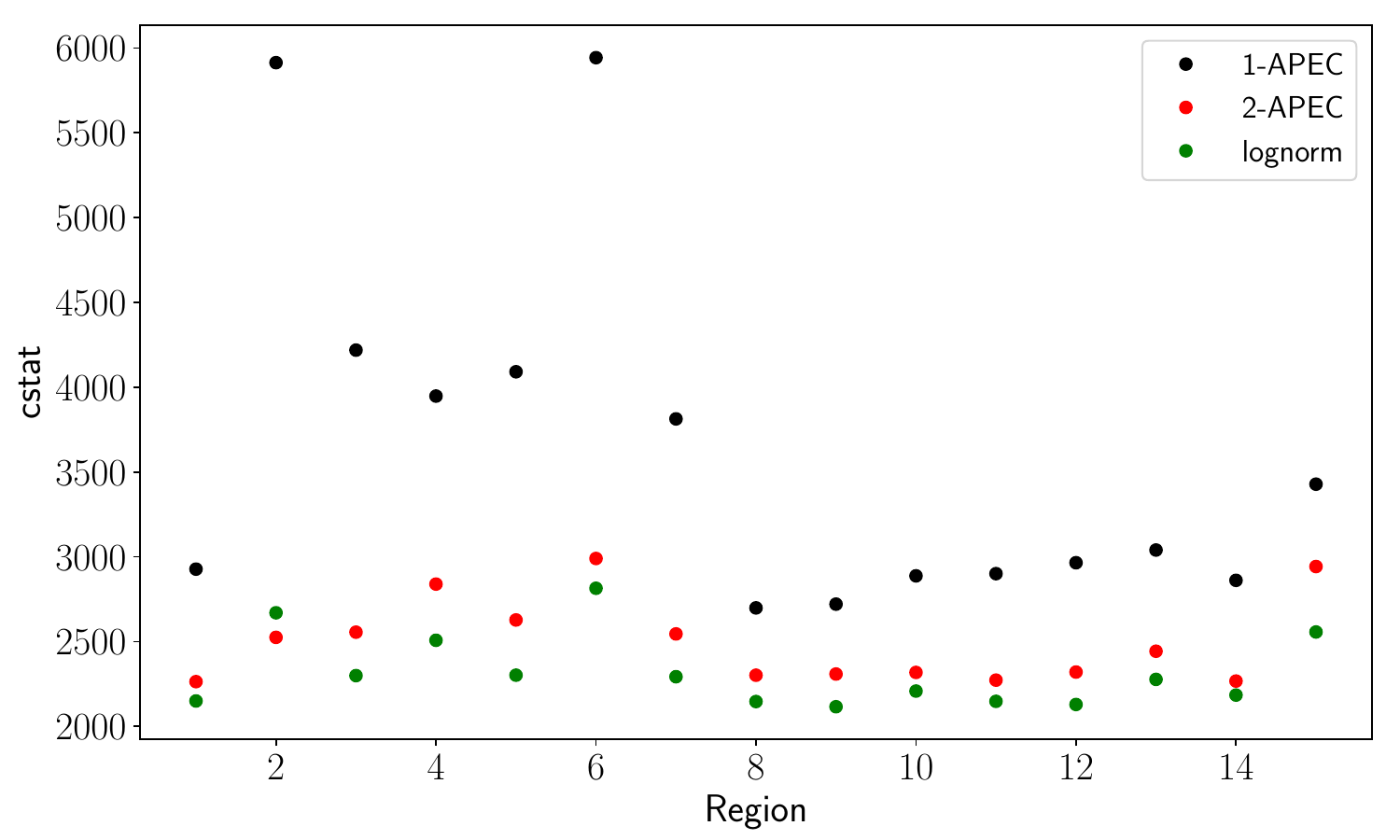} 
\caption{Best-fit cash statistic obtained with the 1-{\tt apec}, 2-{\tt apec} and {\tt lognorm} models.  } \label{cstat_comparison} 
\end{figure}

Following the analysis done in \citet{san20}, we included Cu-$K\alpha$, Cu-$K\beta$, Ni-$K\alpha$, Zn-$K\alpha$, and Al-$\alpha$ instrumental emission lines, and a power-law component with its photon index fixed at 0.136 as background components. 
We also considered the astrophysical background by including a power-law with $\Gamma=1.45$ that accounts for the unresolved population of point sources, one unabsorbed thermal plasma model for the Local Hot Bubble (LHB) emission and one absorbed thermal plasma model ({\tt apec}) for the Galactic halo (GH) emission \citep[e.g.][]{yos09}.
We estimated the temperatures of these components by extracting the spectra from 3 elliptical regions located in the outskirt of the cluster, $\sim 120$~kpc from the cluster center (see Figure~\ref{fig_regions}). 
Table~\ref{tab_bkg} list the best-fit parameters obtained for the CXB, GH, and LHB components. 
Given the proximity with the outskirts of the cluster, we also included the ICM component when modeling the astrophysical background.
Henceforth, the temperatures for these background components were fixed to the best-fit values of $kT_{e}=0.13$~keV and $kT_{e}=0.05$~keV for the fits of the inner regions.

\begin{figure}    
\centering
\includegraphics[width=0.49\textwidth]{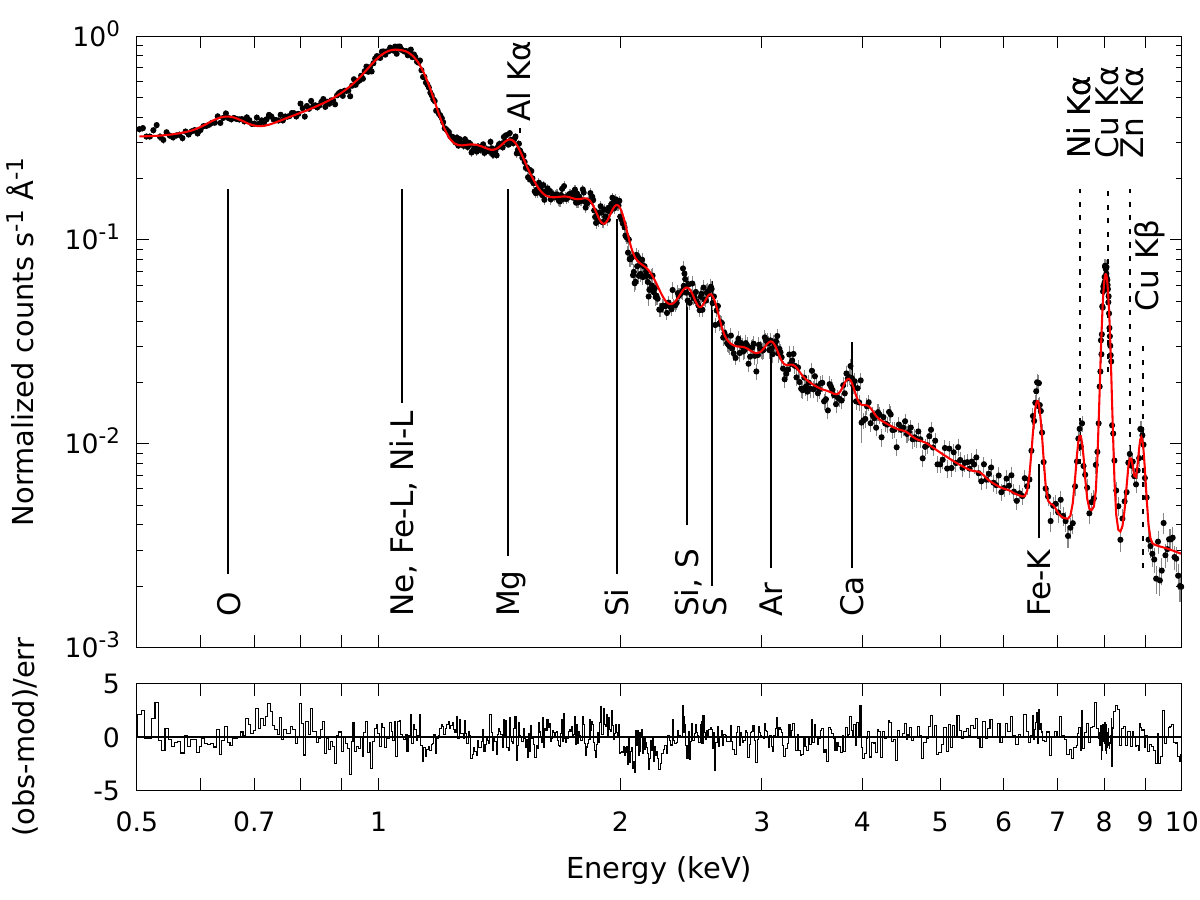} 
\caption{Example spectrum and best-fit model obtained for region 1. The spectrum has been rebinned for illustrative purposes. The line contribution from instrumental background (vertical dashed lines) and from ICM emission (vertical solid lines) are indicated. The lower panel shows the residuals to the fit.} \label{fig_spectra_example} 
\end{figure}

Figure~\ref{fig_spectra_example} shows an example spectrum and residuals from this analysis (i.e. for region 1). 
The solid line indicates the best-fitting {\tt lognorm} model. 
Vertical dashed lines indicate the instrumental Ni K$\alpha$, Cu K$\alpha$,$\beta$, Zn K$\alpha$, and Al K$\alpha$  background lines, included in the model. 
Vertical solid lines indicate the contribution of O, Ne, Mg, Si, S, Ar, Ca, Fe, and Ni to the line emission.
 
 \begin{table}
\caption{Astrophysical background model best-fit parameter. }\label{tab_bkg}
\begin{center}
\begin{tabular}{ccccccc}
\hline
\hline
  & $\Gamma$ & & kT (keV)     \\
\hline  
CXB & 1.45  & GH & $ 0.13 \pm 0.01 $   \\
 &  &  LHB & $0.050\pm 0.007$     \\
\hline 
\end{tabular}
\end{center}
  
\end{table} 

\section{Results }\label{sec_dis}

\subsection{Temperature profile}\label{spec_maps}   
\begin{figure}    
\centering
\includegraphics[width=0.48\textwidth]{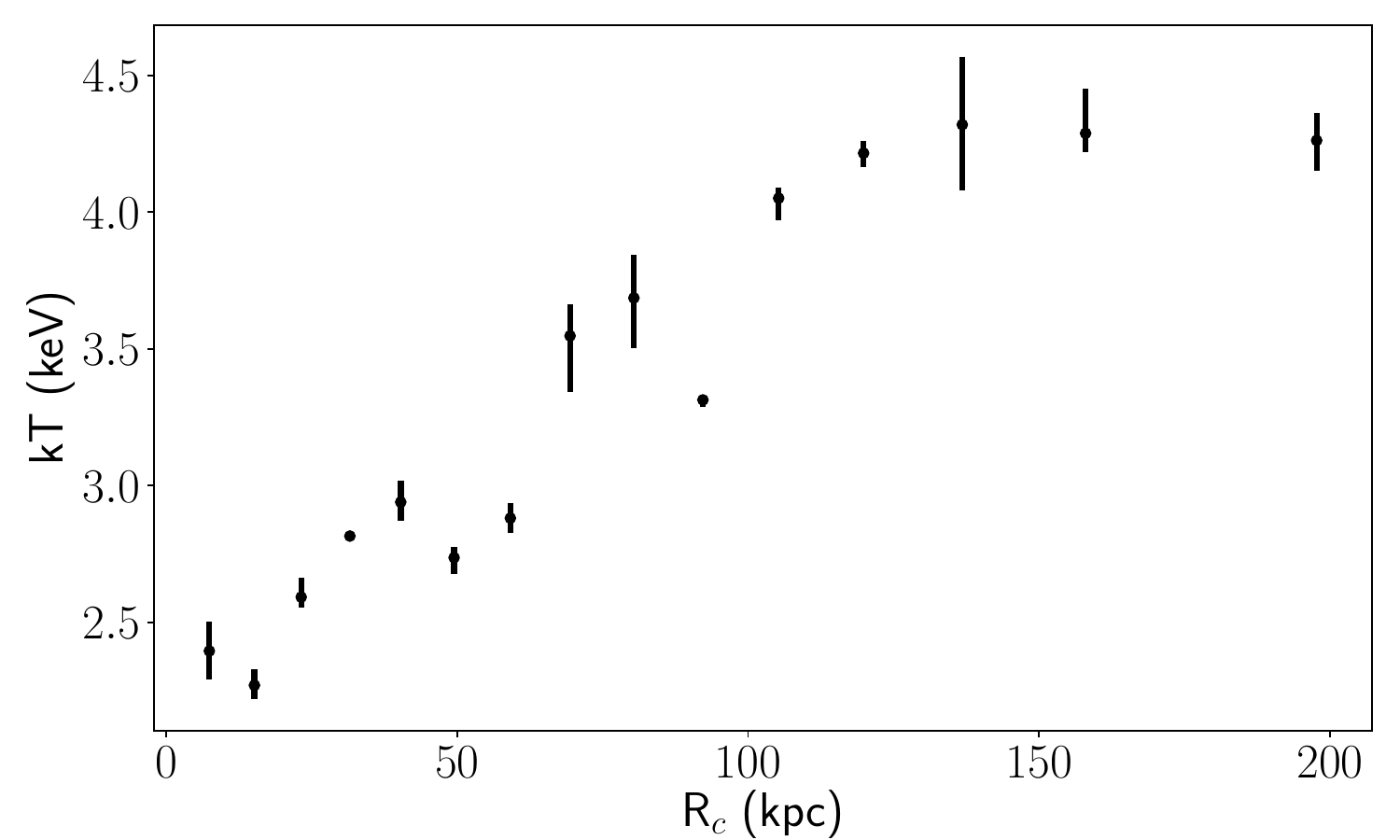}\\
\includegraphics[width=0.48\textwidth]{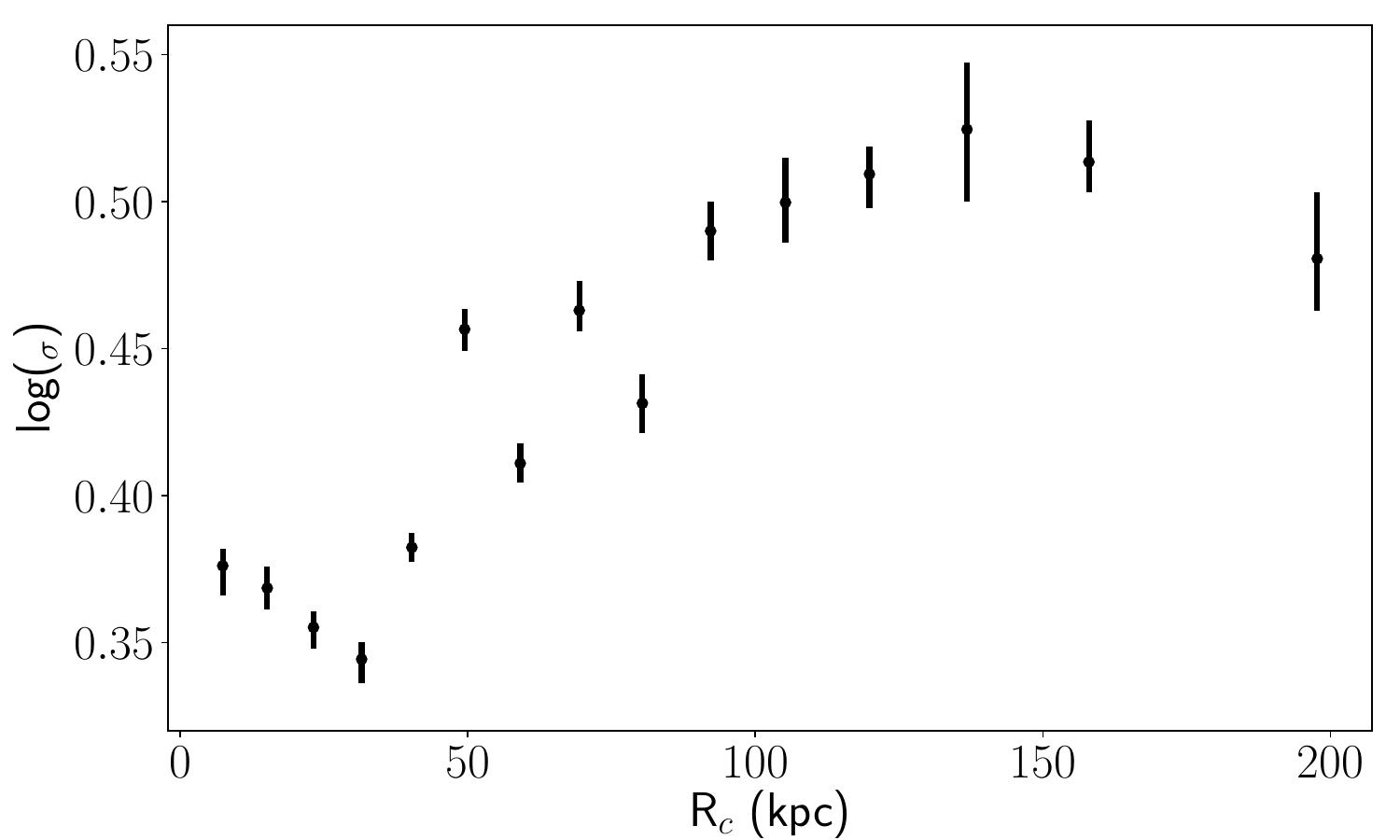} 
\caption{ 
\emph{Top panel:} Temperature profile obtained from the best-fit results. \emph{Bottom panel:} $\log(\sigma)$ profile. 
} \label{fig_kt_sigma} 
\end{figure}

Figure~\ref{fig_kt_sigma} presents the temperature profile (top panel) and $\log(\sigma)$ (bottom panel) derived from the best-fit per region. 
The obtained temperatures generally exhibit lower values than those reported by \citet{gat22b}.
This is expected as our analysis incorporates the soft-energy band. 
The plot distinctly illustrates the apparent discontinuities identified at approximately $\sim 50$ kpc and $\sim 100$ kpc, consistent with the findings of \citet{gat22b}. 
Furthermore, a suggestive indication of a discontinuity at around $\sim 10$ kpc, as reported by \citet{wal13a}, emerges. 
Interestingly, recent studies have demonstrated the presence of turbulence at a driving scale of approximately $\sim 10-20$ kpc in the Centaurus cluster \citep{gat23b}. 
However, these discontinuities which are also observed in the abundance distribution as well as the surface brightness are most likely associated with the cold fronts \citep[see Figures 15-16 in][]{gat22b}.
The distribution of $\log(\sigma)$ emphasizes the necessity of considering a multi-temperature component for all the examined regions (Figure~\ref{fig_kt_sigma}, bottom panel). 
Notably, the $\log(\sigma)$ distribution exhibits the same discontinuities observed in the temperature profile.

\subsection{Velocity profile}\label{spec_maps}   

\begin{figure}    
\centering  
\includegraphics[width=0.48\textwidth]{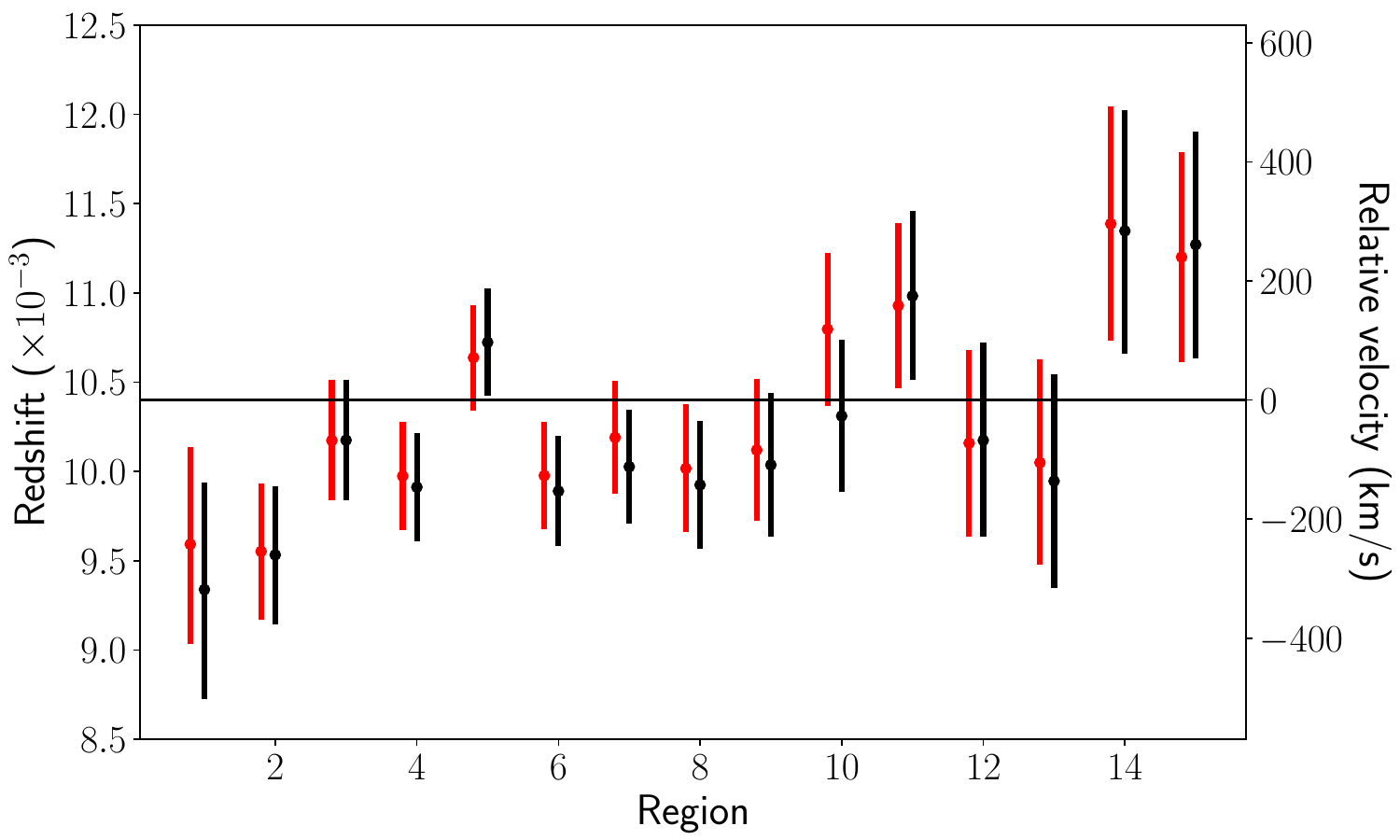}  
\caption{
A comparison between the velocities obtained for each region (black points) and those obtained by \citet[][red points]{gat22b}. The horizontal line indicates the Centaurus cluster redshift. 
}\label{fig_vel} 
\end{figure}

Figure~\ref{fig_vel} illustrates the velocities obtained for each region examined in this study. 
Precise velocity measurements have been achieved with a resolution of $\Delta v\sim 92$ km/s (for ring 6). 
The most significant redshift/blueshift values observed in relation to the Centaurus cluster are $284\pm 204$ km/s (ring 14) and $-318\pm 182$ km/s (ring 1). 
Blueshifted gas appear more frequent as we approach the core of the cluster, although the overall velocity dispersion relative to the system velocity remains low. 
Although the inclusion of the soft-energy X-ray band and the implementation of a multi-temperature model may affect the obtained velocities for the Fe-$K$ complex, Figure~\ref{fig_vel} demonstrates the consistency of our current findings with those reported by \citet{gat22b}, highlighting their robustness.

\subsection{Abundance profiles}\label{circle_rings} 
\begin{figure*}    
\centering
\includegraphics[width=0.33\textwidth]{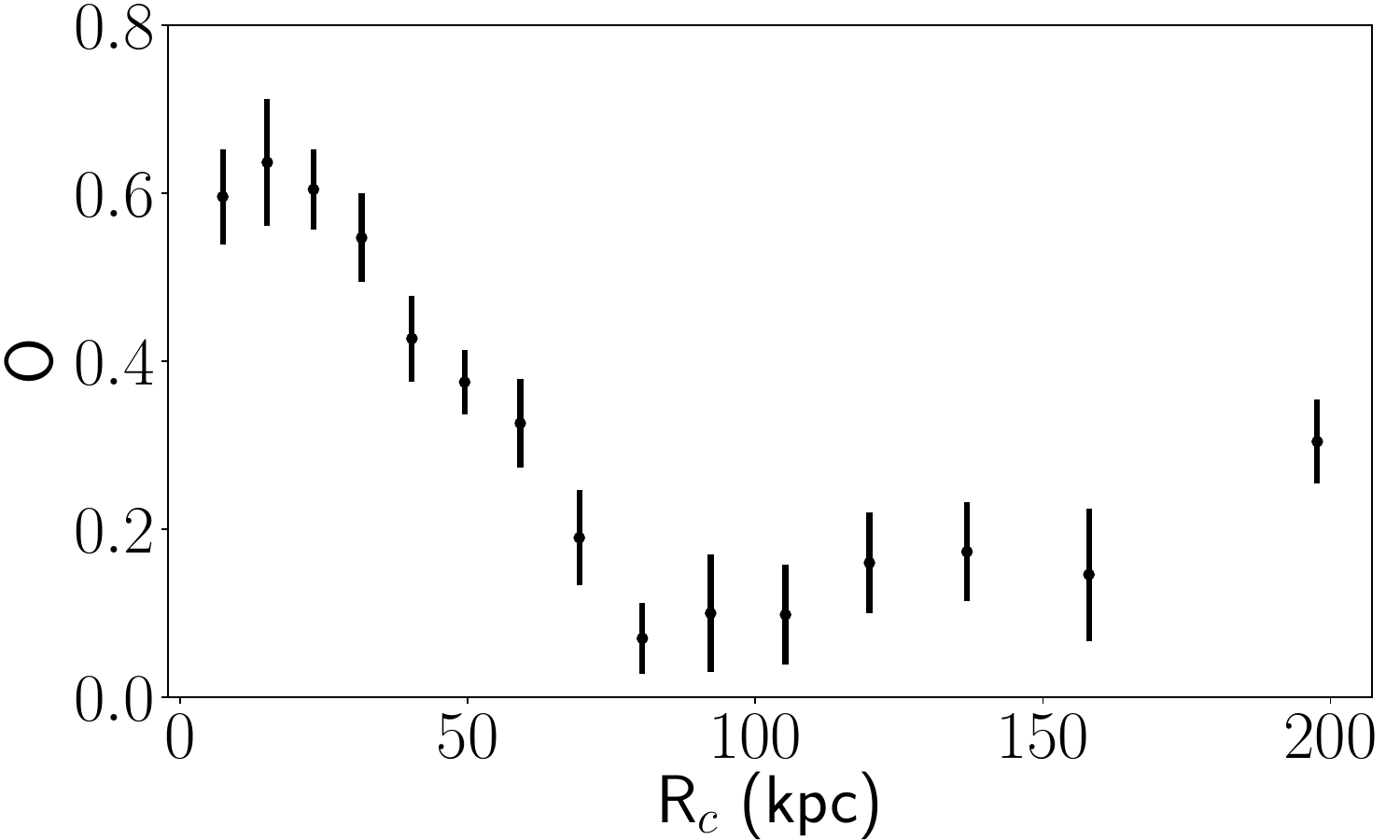} 
\includegraphics[width=0.33\textwidth]{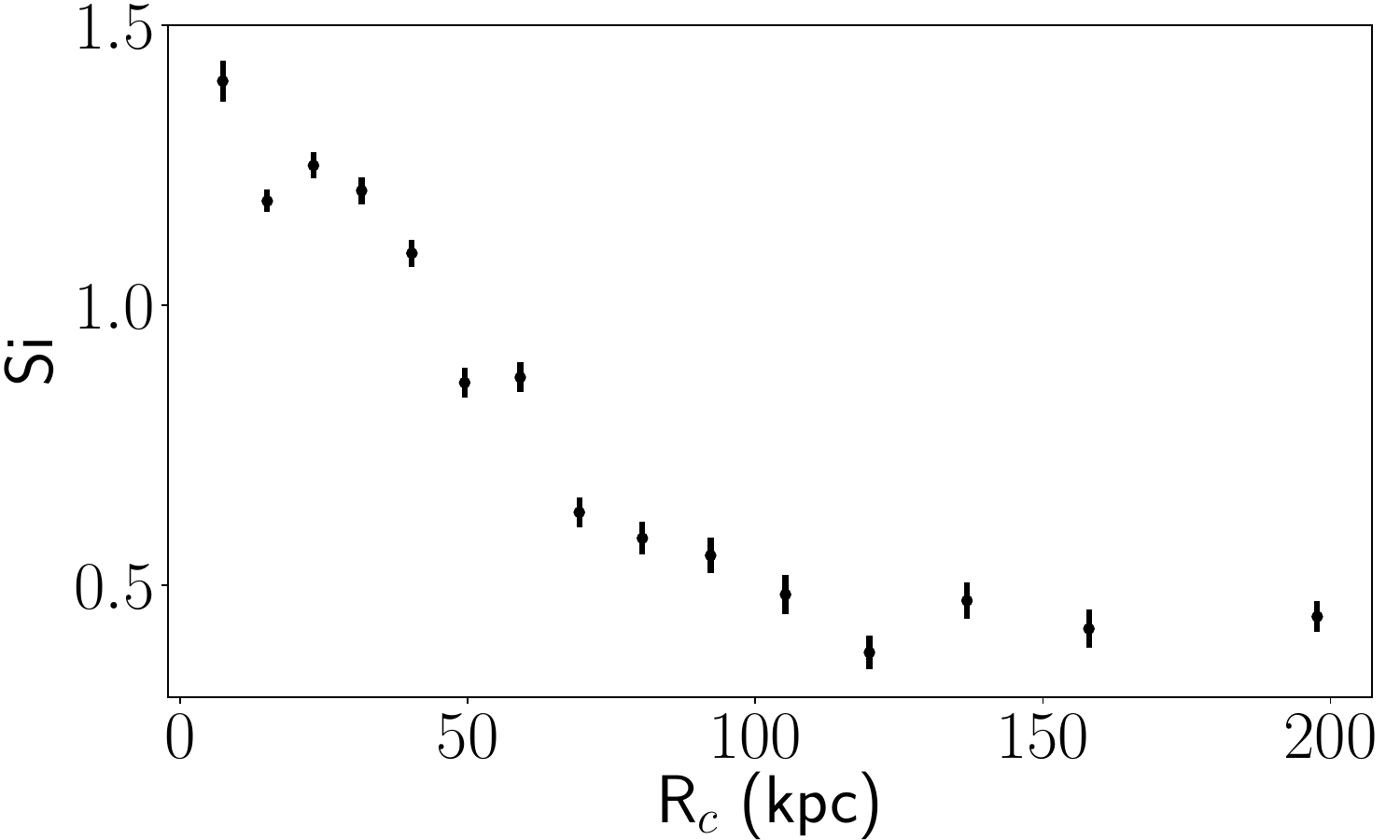}
\includegraphics[width=0.33\textwidth]{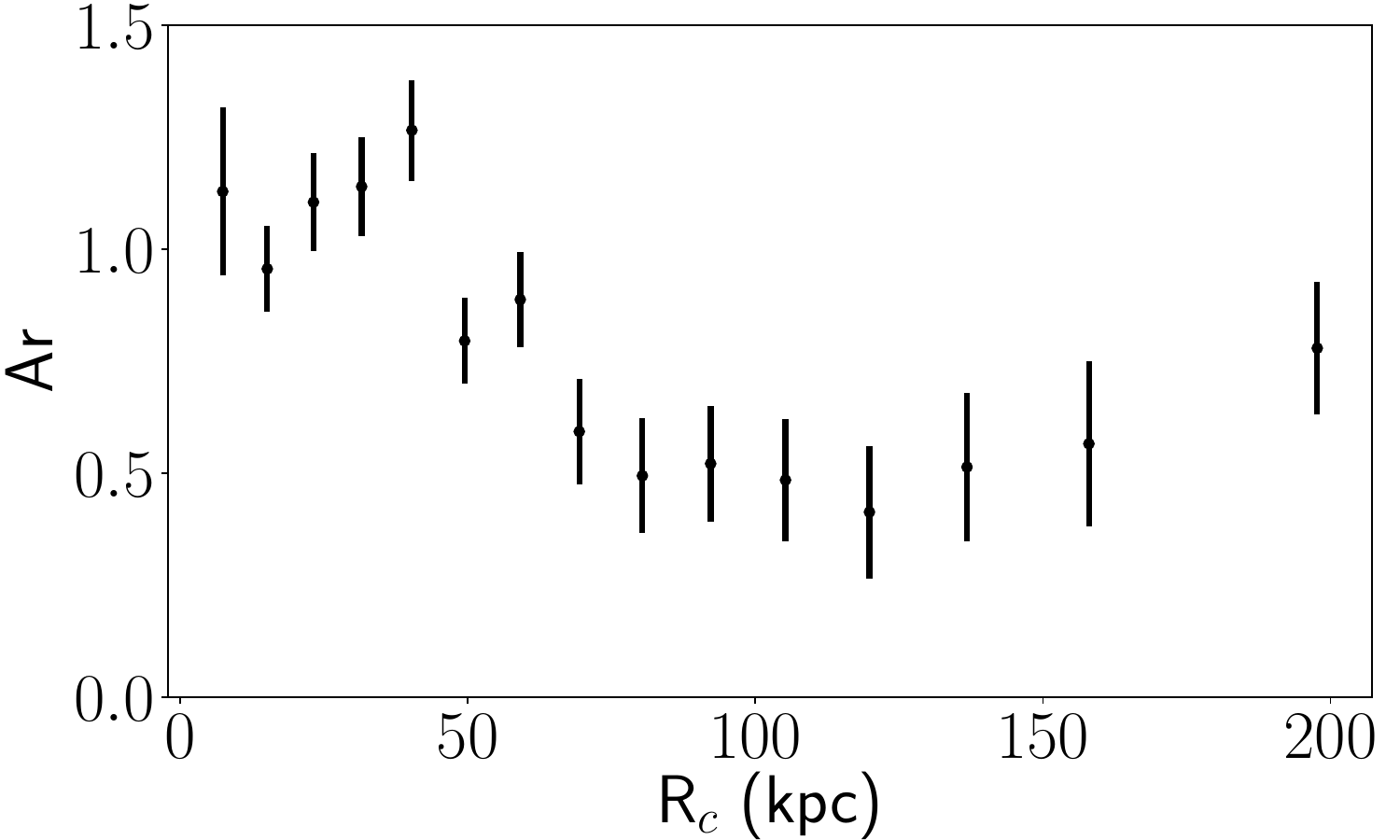}\\
\includegraphics[width=0.33\textwidth]{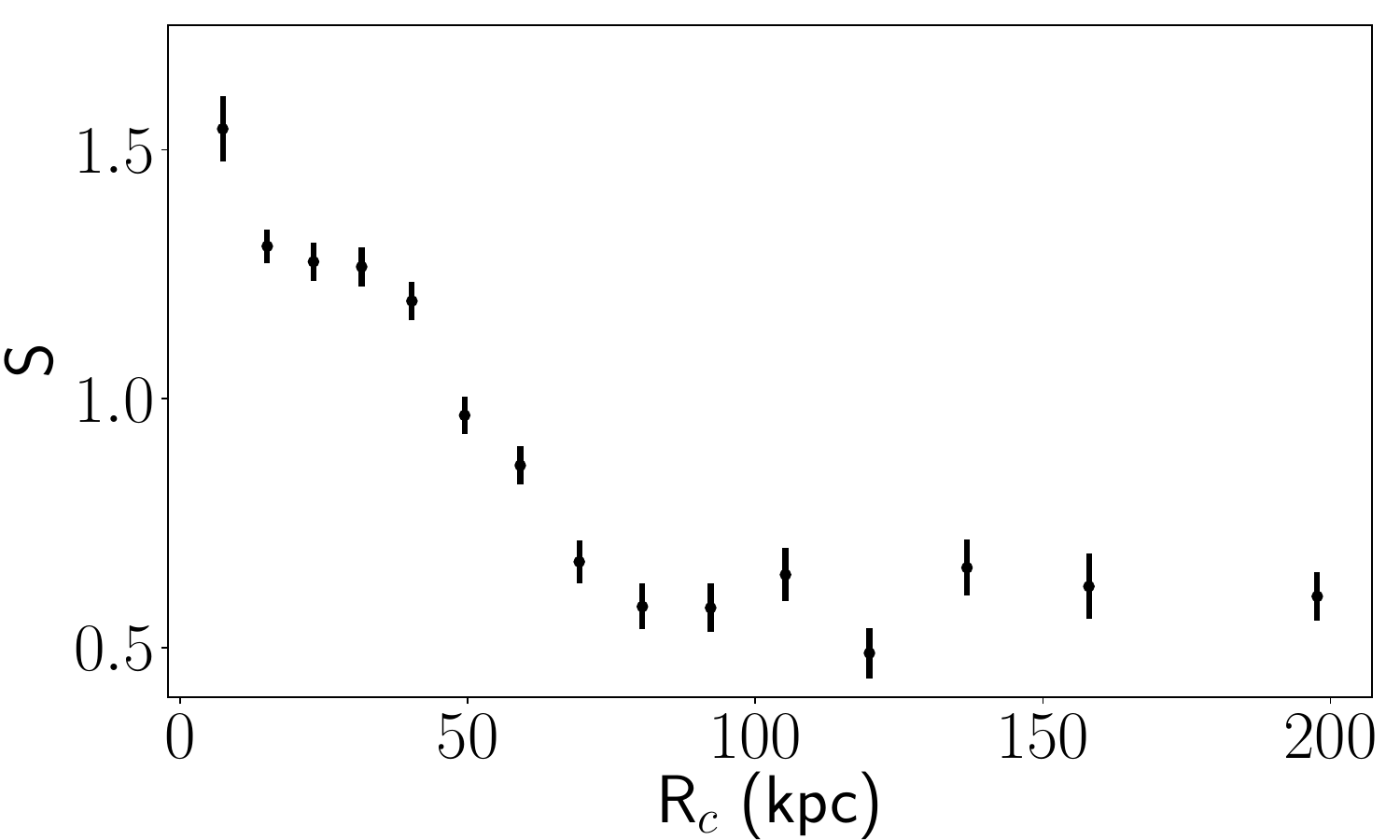}
\includegraphics[width=0.33\textwidth]{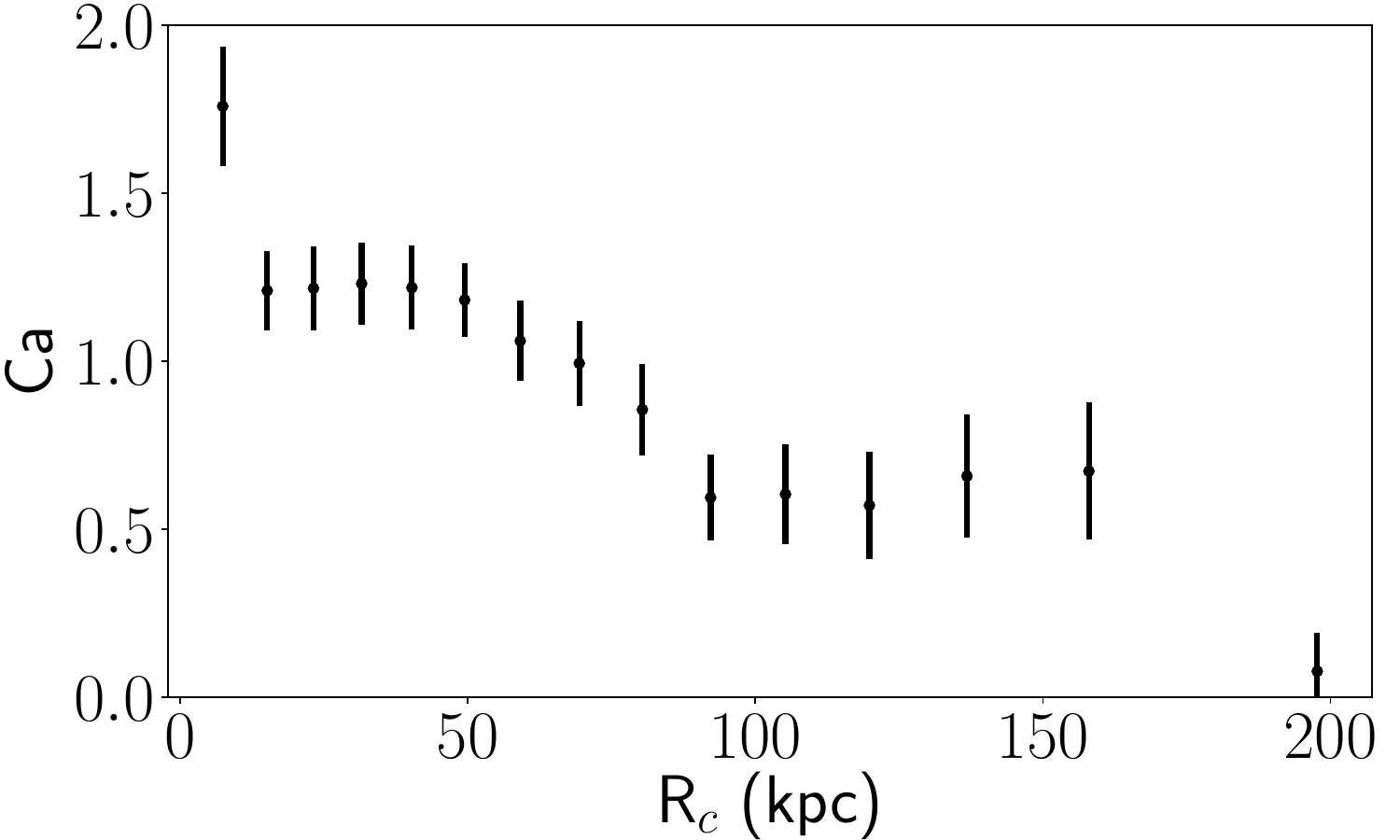} 
\includegraphics[width=0.33\textwidth]{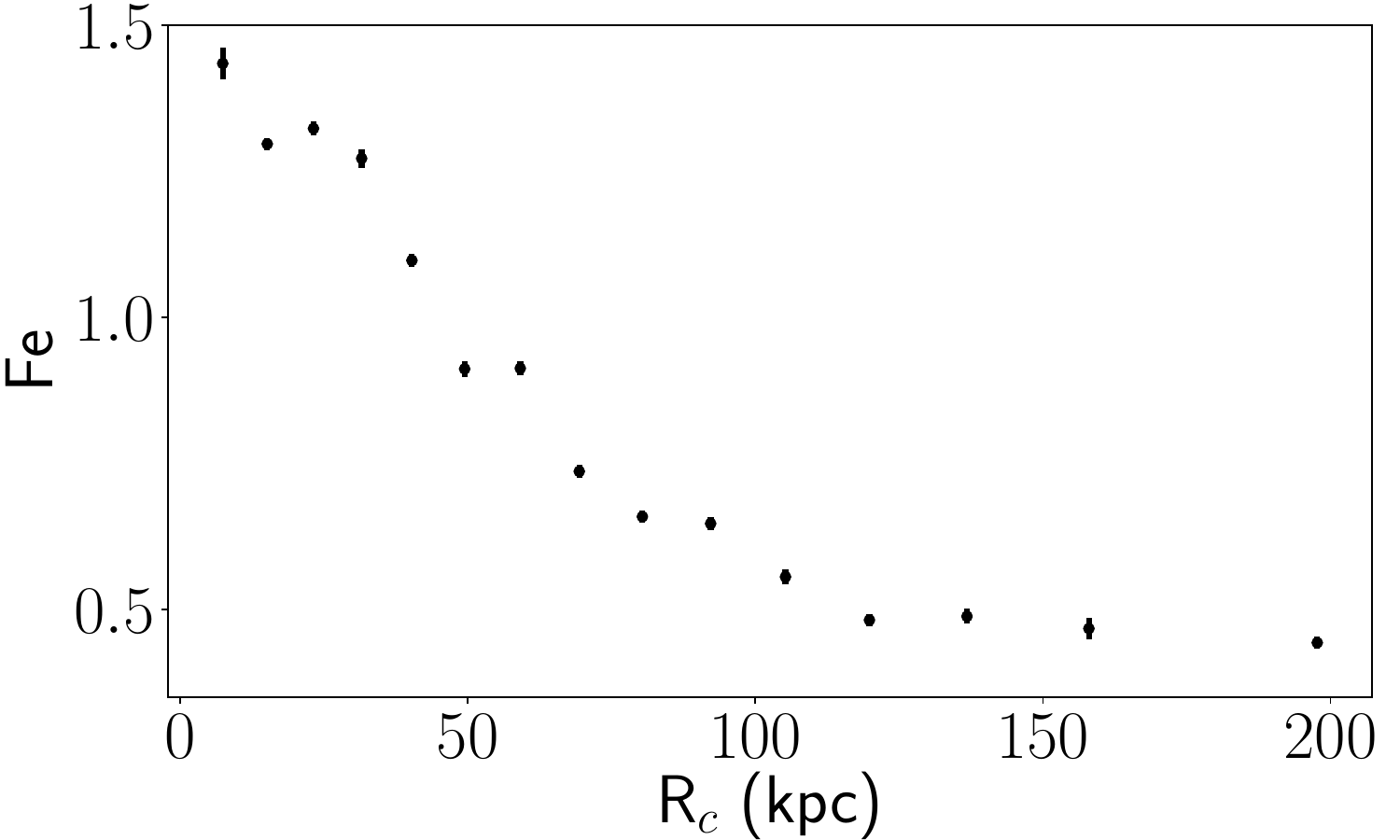}  
\caption{
Abundance profiles obtained from the best-fit results.
} \label{fig_abund_all} 
\end{figure*}

Figure~\ref{fig_abund_all} presents the elemental abundances derived from the best-fit analysis per region relative to the solar values. The results reveal prominent negative gradients in the O, Si, S, and Fe abundances for distances $<90$ kpc, which then transition to flatter profiles. The Ca abundance, on the other hand, exhibits a relatively constant profile for shorter distances. Notably, discontinuities in the abundance profiles align with those identified in the temperature distribution at approximately $\sim 15$ kpc, $\sim 50$ kpc and $\sim 100$ kpc (see Figure~\ref{fig_kt_sigma}). Furthermore, comparing the temperature distribution highlights that the cooler gas is more iron-rich. 

\begin{figure}    
\centering
\includegraphics[width=0.46\textwidth]{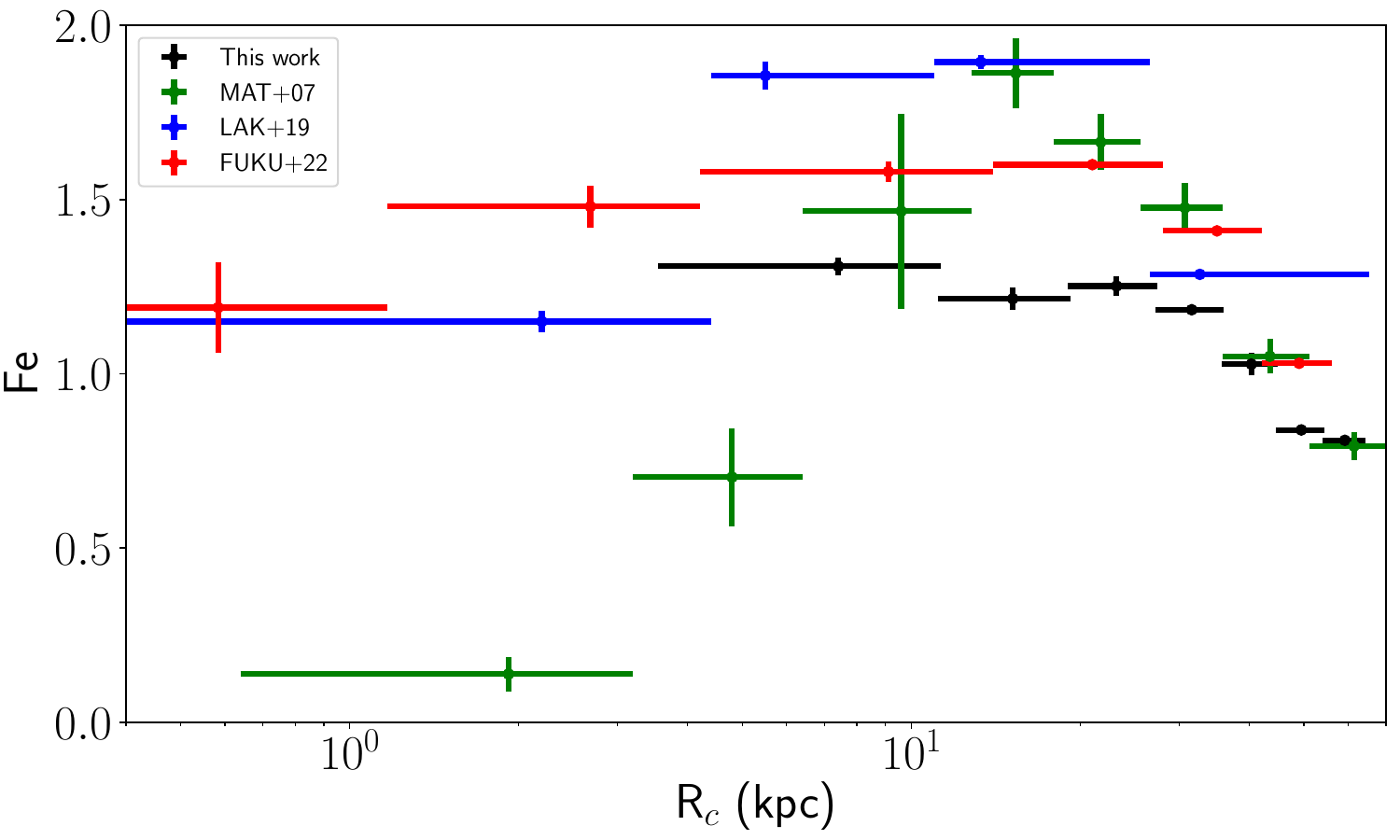}  
\caption{
Fe abundance distribution near the cluster core. Black points correspond to the results obtained from the best-fit results. Green points correspond to \citet{mat07}, blue points correspond to \citet{lak19} and red points correspond to \citet{fuk22}.
} \label{fig_fe_comp} 
\end{figure}

Figure~\ref{fig_fe_comp} shows a comparison between the values derived from the best-fit and those obtained from previous analysis of {\it XMM-Newton} EPIC-pn observations by \citet[][referred to as MAT+07]{mat07}, \citet[][referred to as LAK+19]{lak19}, and \citet[][referred to as FUKU+22]{fuk22}. The abundances provided by MAT+07 and LAK+19 have been rescaled to \citet{lod09}. Notably, the decline in iron (Fe) abundance reported by LAK+19 and FUKU+22 is particularly pronounced for distances  $<8$~kpc, a region not covered in our analysis. Moreover, the plot indicates a significant discrepancy between the abundances obtained by LAK+19 and FUKU+22, with the former being $\sim 1.3$ times higher than the latter. The curve is relatively flat before the decrease, similar to our findings from the best-fit results. In contrast, MAT+07 exhibits a much sharper decrease in abundance from larger distances ($<15$~kpc). However, they fit the X-ray spectra with the {\tt mekal} model \citep{lie95}. In that sense, changes in the atomic data involved may explain differences between the {\tt mekal} and {\tt apec} models. We have also found differences in the abundance profiles when comparing with previous measurements obtained with {\it Chandra} observations. Notably, the drop in metallicity was identified within distances of $< 10$~kpc from the cluster center \citep{san02,san06b,pan13,san16}.

\subsection{ICM chemical enrichment from SN}\label{sec_snr} 

\begin{figure*}    
\centering
\includegraphics[width=0.33\textwidth]{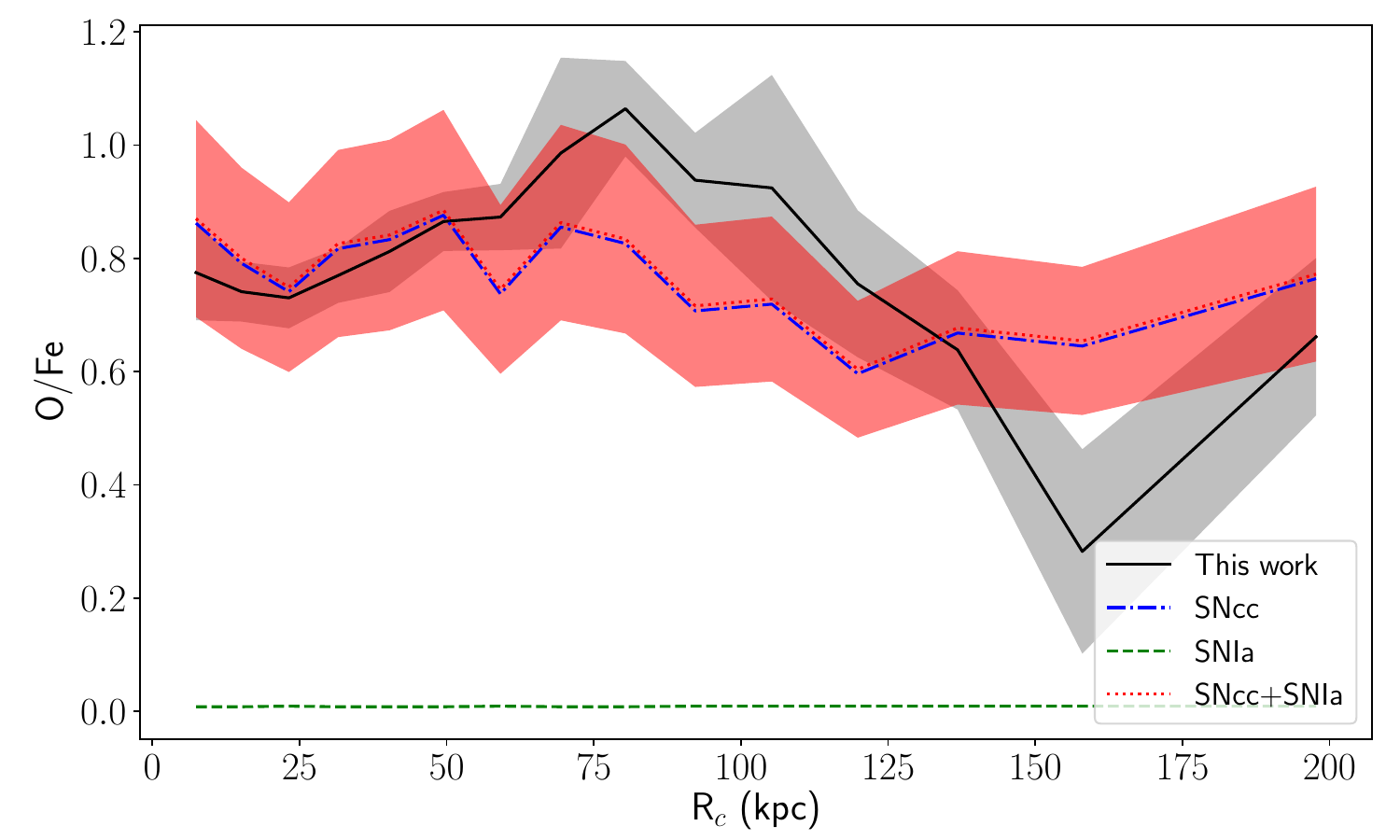} 
\includegraphics[width=0.33\textwidth]{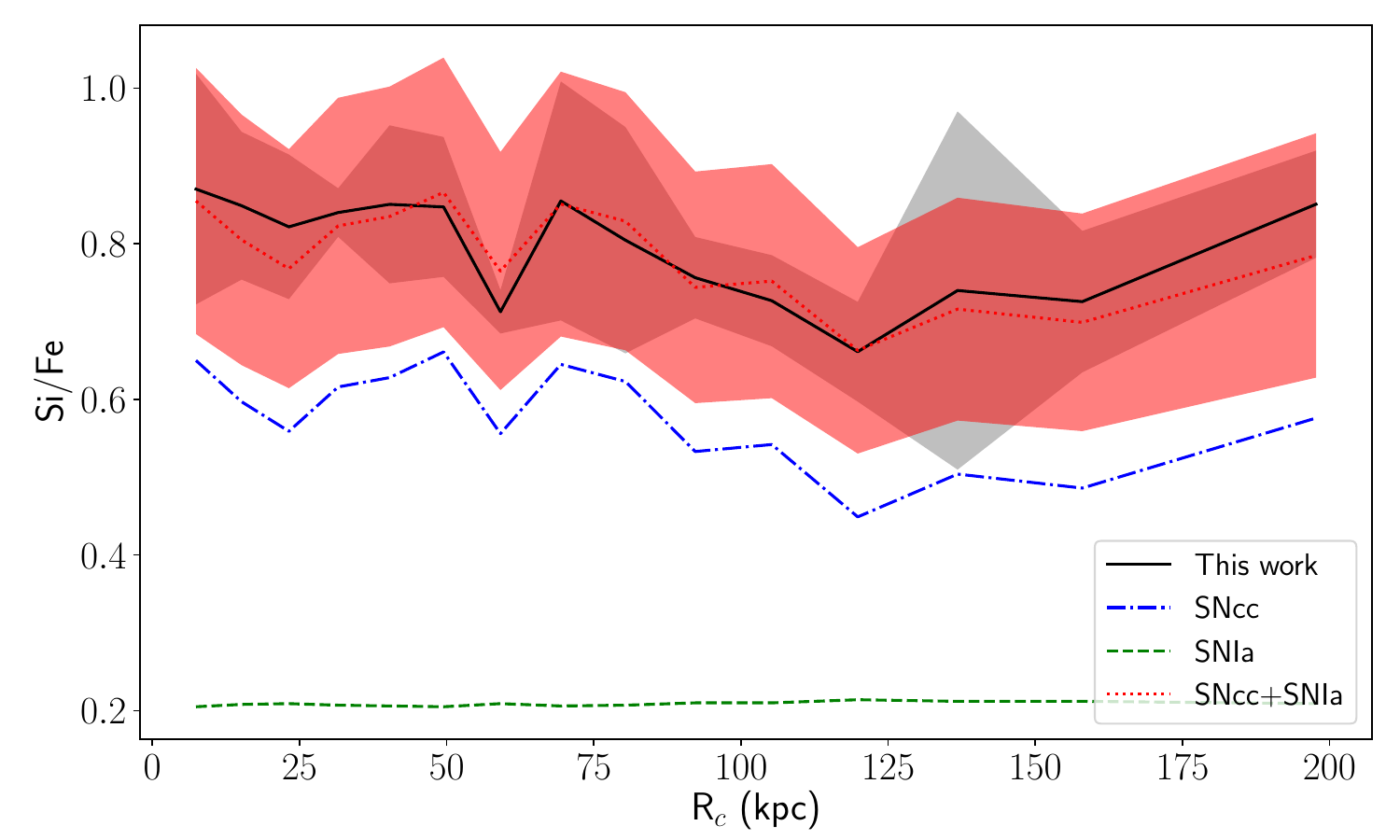} 
\includegraphics[width=0.33\textwidth]{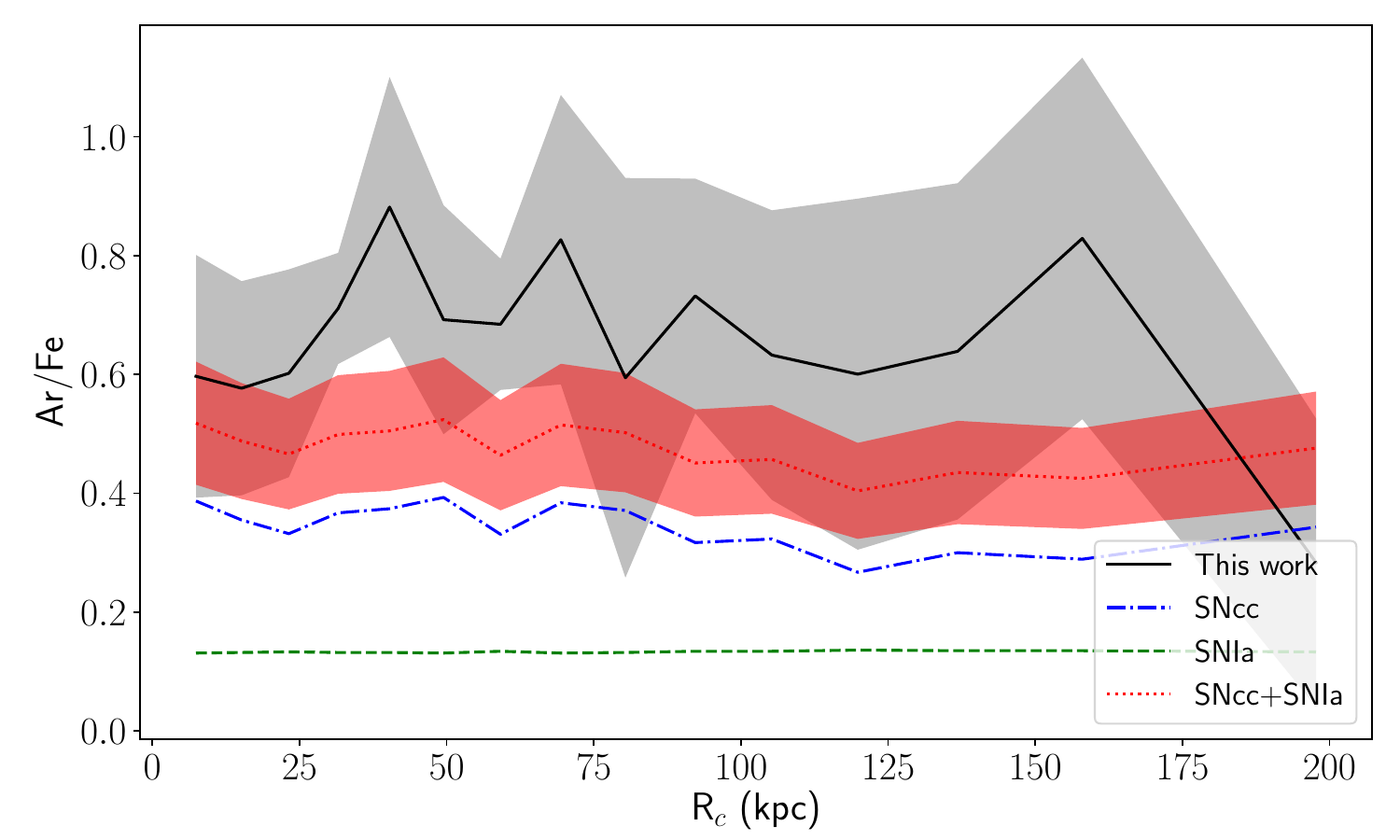}\\
\includegraphics[width=0.33\textwidth]{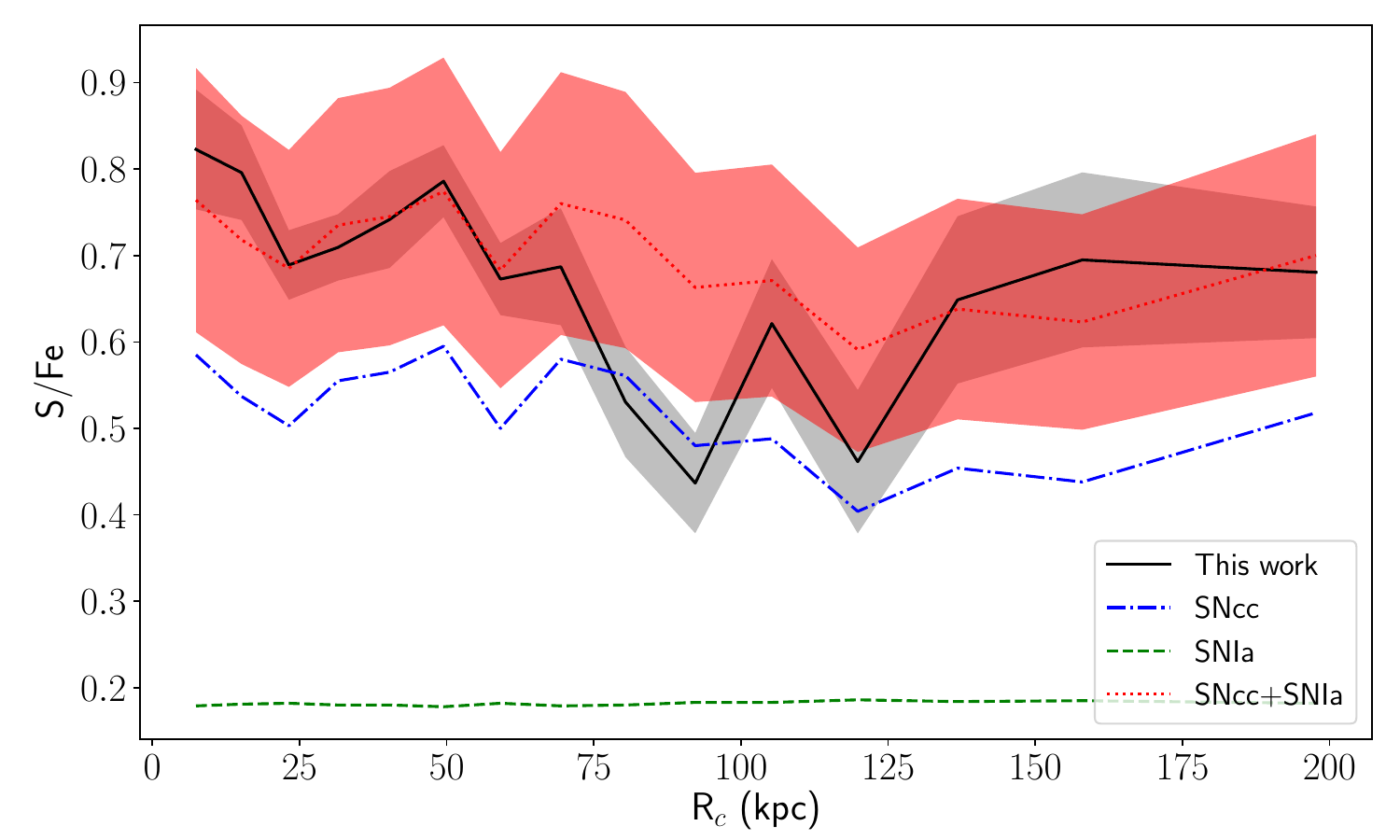} 
\includegraphics[width=0.33\textwidth]{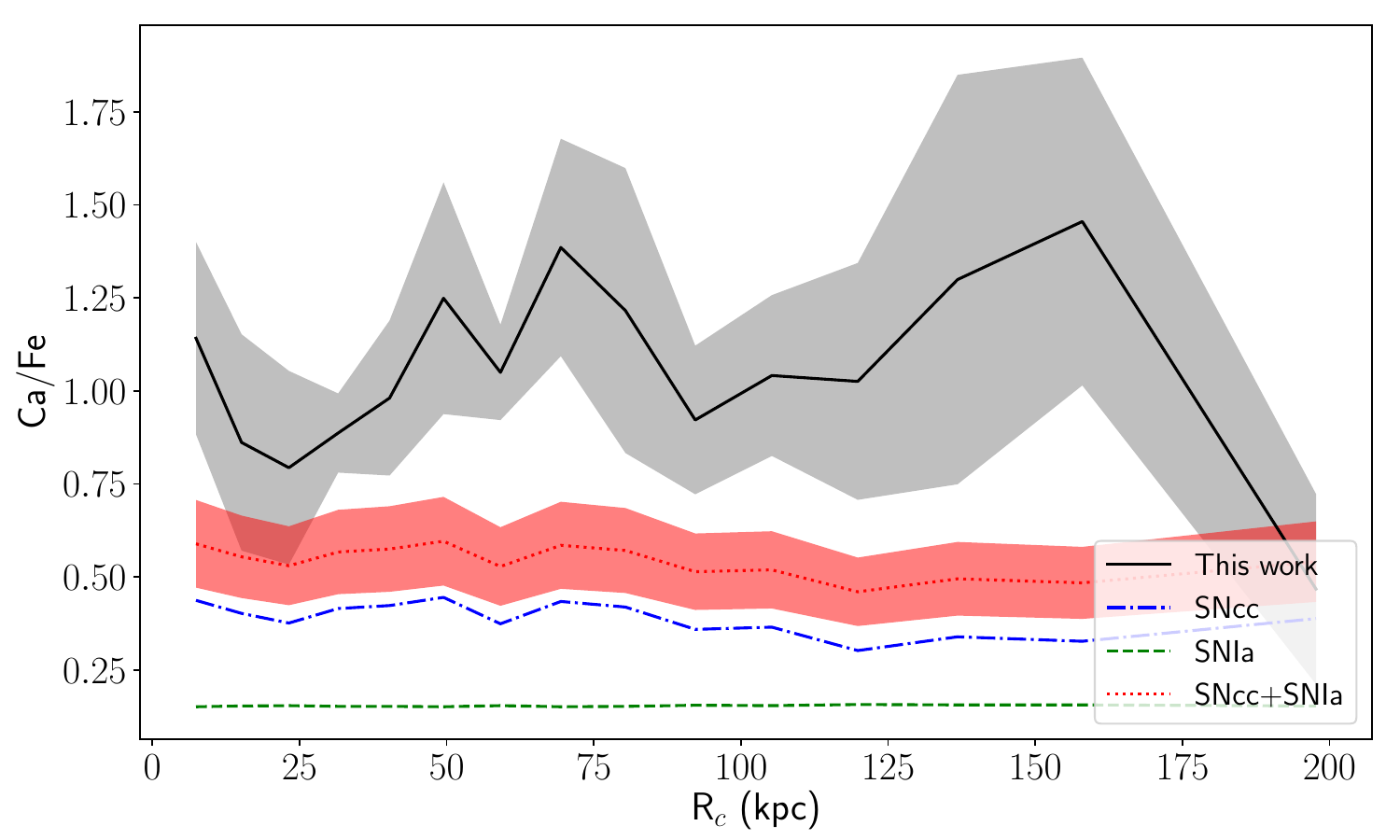} 
\caption{
Abundance ratio profiles, relative to Fe. The gray shaded areas indicate the mean values and the 1$\sigma$ errors. The SNcc (blue line), SNIa (green line) contribution to the total SN ratio (red shaded area) from the best fit model are included (see Section~\ref{sec_snr}). 
 } \label{fig_ratios} 
\end{figure*}

Figure~\ref{fig_ratios} shows the X/Fe ratio profiles for all elements measured (gray-shaded regions). The Si/Fe, S/Fe, and Ca/Fe ratios are close to the solar values for all distances. 
On the other hand, the O/Fe and Ar/Fe tend to be lower than solar.  
We noted that both Ar/Fe and Ca/Fe profile shapes are similar even for large distances. 
However, \citet{fuk22} show discrepancies of about 50$\%$ between CCD detectors in the Ca/Fe ratios. 

To model the contribution from different SN yield models to these abundance ratio profiles, we used the {\tt SNeRatio} python code \citep{erd21}. 
Once a set of ICM abundances is defined, the model fits it with a combination of multiple progenitor yield models to calculate the relative contribution that better provides the data. 
For the SNcc yields, we included models from \citet{nom13} with initial metallicity values of Z $=$ 0.0, 0.001, 0.004, 0.008, 0.02, 0.05. 
The SNcc yields were integrated with Salpeter IMF over the mass range of 10-70 M$_\odot$. 
For SNIa, we considered a set of 3D SNIa models near Chandrasekhar-mass, including two different explosion mechanisms: delayed detonation from \citet{sei13} and pure deflagration from \citet{fin14}. 
Also, we assumed the same SNe model along the radii. 
Such a combination of models has been used in recent enrichment studies \citep{mer17,sim19,mer20,gat23a}.

\begin{figure}    
\centering
\includegraphics[width=0.48\textwidth]{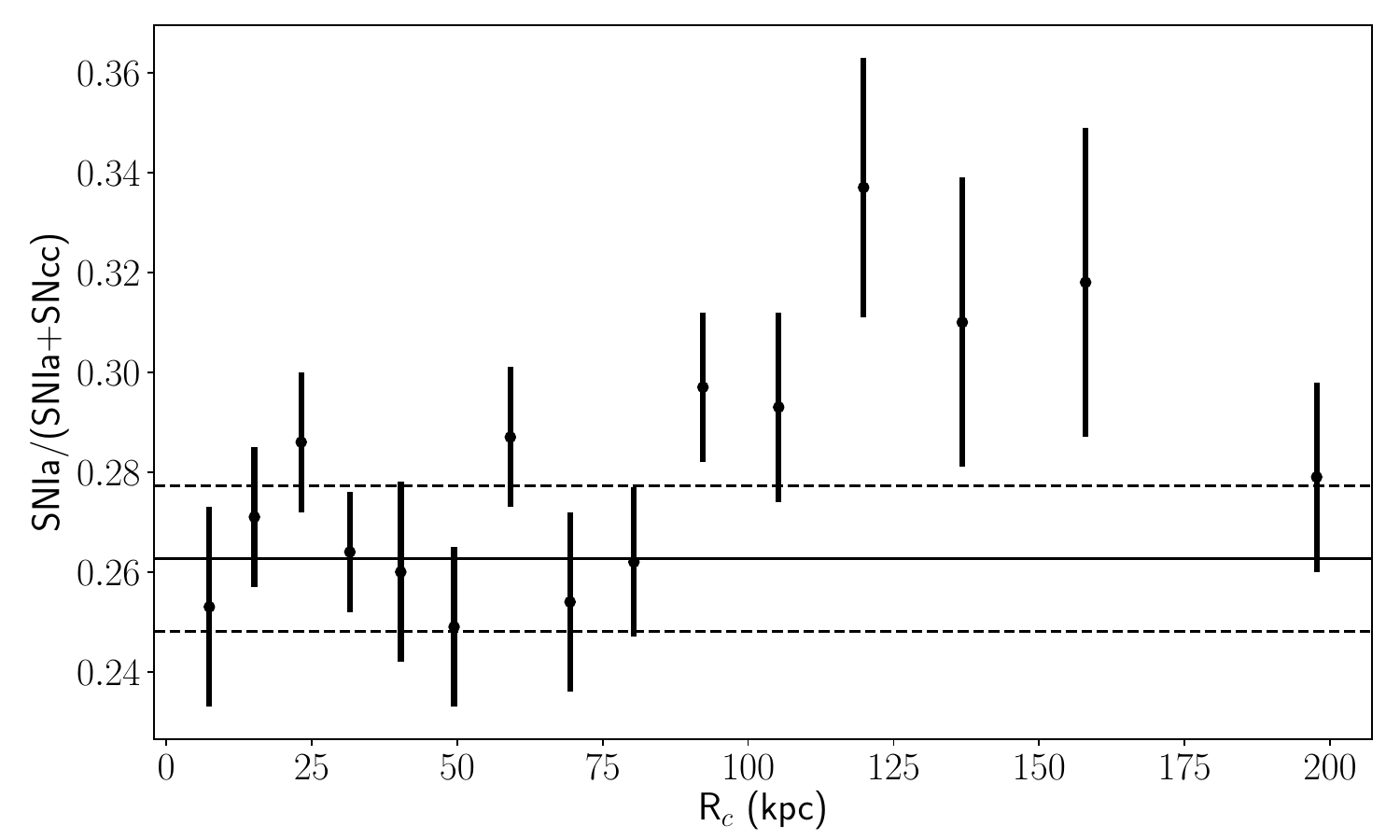} 
\caption{
SNIa contribution to the total chemical enrichment as function of the distance. The model includes O, Si, S, Ar, Ca and Fe abundances. The black line represents 1$\sigma$ confidence interval of constant fit.
} \label{fig_snia_distribution} 
\end{figure}

We determine the best linear combination of SNIa and SNcc models that better fit the data by minimizing the sum of their $\chi^{2}$ values in quadrature. 
The best-fit model corresponds to an initial metallicity Z$=0.0004$ for SNcc and a delayed detonation 3D N10 model for SNIa \citep[see Table~1 in][]{sei13}. 
The model is included in Figure~\ref{fig_ratios}.  
Table~\ref{tab_snr_contribution} and Figure~\ref{fig_snia_distribution} show the SNIa contribution to the total enrichment by this set of models as a function of the distance to the cluster center. 
The O/Fe, Si/Fe, and S/Fe ratios are better reproduced than the Ar/Fe and Ca/Fe ratios.
The SNIa contribution required is $<40\%$ for all radii, similar to the model obtained for the Virgo cluster \citep{gat23a}. 
A line fit gives almost zero slope (2.01$\times 10^{4}$) and shows that the fractional SNIa contribution to the total SNe tends to be constant, with a value of 0.262$\pm$0.014 with $\chi^{2}/d.o.f.=1.24$ (see Figure~\ref{fig_snia_distribution}).

\begin{table}
%\scriptsize 
\caption{\label{tab_snr_contribution}SNIa contributions to the total chemical enrichment. }
\centering
\begin{tabular}{ccccccc}
\\
Radius  & SNIa & Radius & SNIa \\
 (kpc) & & (kpc)   & \\ 
\hline 
7.41 & $25\pm 2 \% $ &80.37& $26\pm 1 \% $\\ 
15.16 & $27\pm 1 \% $&92.23& $29\pm 1 \% $\\ 
23.19 &$28\pm 1 \% $ &105.25&$29\pm 2 \% $ \\
31.56&$26\pm 1 \% $&119.84&$33\pm 3 \% $ \\ 
40.30 &$26\pm 2 \% $ &136.80& $31\pm 3 \% $\\ 
49.47 &$25\pm 2 \% $ &157.99& $32\pm 3 \% $\\
59.51& $29\pm 1 \% $&197.69&$28\pm 2 \% $ \\
69.40& $25\pm 2 \% $\\
 \hline
\end{tabular}
\end{table}

\section{Discussion}\label{sec_dis2} 

\subsection{Metallicity drop}\label{sec_dis2_drop}
The origin of the drop in metallicity found in some galaxy clusters is still a subject of debate. 
It has been shown that effects from resonant scattering contribute to only a negligible fraction of the missing Fe \citep{san06c,gen17}. 
Another proposed explanation is the dust depletion of metals in the very core of the ICM \citep{pan15,mer17,lak19}. 
However, due to systematic and statistical uncertainties in the abundance measurements, it is challenging to firmly confirm or rule out the dust depletion scenario \citep{mer17,fuk22}. 
Multiple reasons may explain the absence of the metallicity drop in our {\it XMM-Newton} analysis. 
First, the multi-temperature modeling of the spectra may affect the abundance measurements, particularly in clusters with temperatures in the range of $3-4$~keV and near the cool core, that is before the plasma became progressively isothermal \citep[e.g., as observed in the analysis of the Abell 2028 cluster by][]{gas10}. 
Second, calibration discrepancies between the {\it XMM-Newton} and {\it Chandra} observatories have been reported in previous works \citep{sch15,mad17}, although these discrepancies primarily impact the shape of the continuum as well as the Fe L complex. 
Additionally, due to the large point-spread function (PSF) of the EPIC-pn camera, we have analyzed relatively larger bin sizes for the innermost region of the cluster, preventing us from fully resolving such a metallicity drop. Figure~\ref{fig_fe_comp} compares the iron abundance found in this analysis and results from \citep{mat07,lak19,fuk22}. The figure clearly shows the differences between the multiple measures.
In this sense, \citet{fuk22} suggested that the abundance drop might be influenced, in part, by potential systematic uncertainties in the response matrices and atomic data utilized in their analysis of both {\it Chandra} and {\it XMM-Newton} observations of the Centaurus cluster. However, any analysis for RGS cross-dispersion slices within $<5$ arcsec seems overly optimistic considering XMM-Newton point-spread PSF.

\subsection{Ar/Fe and Ca/Fe abundance ratios}\label{sec_dis2_ratios}
Regarding the underestimation of Ca/Fe and Ar/Fe abundance ratios from the SNe models (see Figure~\ref{fig_ratios}), similar results have been found from previous CCD measurements \citep{dep07,mer16,sim19}. To explain such deficit in theoretical supernova models compared with XMM-Newton measurements, \citet{dep07} found an improvement when using an empirically modified delayed detonation model, which uses Tycho type Ia supernova remnant as a calibration source. Also, calcium-rich gap transients, a subclass of SNia, may contribute to the ICM enrichment \citep{mul14}. However, both models tend to overproduce Cr or decrease Ar production. A comparison with ICM abundances obtained with future high-resolution spectroscopy instruments will help to understand the reliability of CCD spectra measurements better.

\subsection{SNIa fraction}\label{sec_dis2_snia} 
The uniformity in the SNIa percentage contribution to the total SNe (see Figure~\ref{fig_snia_distribution}) is consistent with an early enrichment of the ICM scenario. That is, chemical elements expelled from the galaxies during the protocluster phase were mixed and deposited into the ICM environment, probably during the period of maximal star formation and black hole activity, and before the ICM became stratiﬁed \citep{wer13,man17,sim19}. Otherwise, we should expect a strong peak of SN Ia towards the central regions, which is not observed.
It is worth mentioning that further effort in improving theoretical models of supernova nucleosynthesis is essential to reduce uncertainties in the yield calculations (e.g., including neutrino physics). Finally, \citet{mer17} concluded that sudden changes in the SNIa contribution at the outskirts cannot be entirely excluded even when accounting for various systematic uncertainties.

\section{Conclusions and summary}\label{sec_con} 
We have analyzed {\it XMM-Newton} EPIC-pn observations of the Centaurus cluster to study the radial profiles of O, Si, S, Ar, Ca and Fe in the ICM. 
In this Section, we briefly summarize our findings. Our main findings and conclusions are: 
 
\begin{enumerate}
\item We found velocities in good agreement with \citet{gat22b}, despite the inclusion of the soft energy band (i.e. $<4$~keV). 
The blueshifted gas tends to have a lower temperature than the redshifted gas.
\item The temperature and abundances profile show discontinuities around $\sim15$~kpc, $\sim50$~kpc and $\sim100$~kpc. 
The latter could be associated with cold fronts while the former is of the same order of magnitude as the driving scale of turbulence for the cluster. 
We found that the cooler gas is more metal-rich.
\item We studied the contribution from different SN yield models to the X/Fe ratio profiles elements O, Si, S, Ar and Ca. 
The best-fit model corresponds to an initial metallicity Z$=0.0004$ for SNcc and a delayed detonation 3D model for SNIa which roughly reproduces the observed abundance patterns of O/Fe, Si/Fe, and S/Fe ratios. 
For this model, the SNIa ratio over the total cluster enrichment tends to be uniform and requires an SNIa contribution for almost all radii ($<40\%$). 
Such uniformity in the SNIa percentage contribution supports an early enrichment of the ICM scenario, with most of the metals present being produced before clustering. 
 
\end{enumerate}

Future work includes a detailed analysis of the 2D spatial distribution of elemental abundances to study the morphological features (e.g., clumpiness) and the interaction between the AGN and the surrounding environment.

\section{Acknowledgements} 
This work was supported by the Deutsche Zentrum f\"ur Luft- und Raumfahrt (DLR) under the Verbundforschung programme (Messung von Schwapp-, Verschmelzungs- und R\"uckkopplungsgeschwindigkeiten in Galaxienhaufen). This work is based on observations obtained with XMM-Newton, an ESA science mission with instruments and contributions directly funded by ESA Member States and NASA. This research was carried out on the High Performance Computing resources of the cobra cluster at the Max Planck Computing and Data Facility (MPCDF) in Garching operated by the Max Planck Society (MPG).
\subsection*{Data availability}
The observations analyzed in this article are available in the {\it XMM-Newton} Science Archive (XSA\footnote{\url{http://xmm.esac.esa.int/xsa/}}).

\bibliographystyle{mnras}

\newcommand{\noop}[1]{}
\begin{thebibliography}{}
\makeatletter
\relax
\def\mn@urlcharsother{\let\do\@makeother \do\$\do\&\do\#\do\^\do\_\do\%\do\~}
\def\mn@doi{\begingroup\mn@urlcharsother \@ifnextchar [ {\mn@doi@}
  {\mn@doi@[]}}
\def\mn@doi@[#1]#2{\def\@tempa{#1}\ifx\@tempa\@empty \href
  {http://dx.doi.org/#2} {doi:#2}\else \href {http://dx.doi.org/#2} {#1}\fi
  \endgroup}
\def\mn@eprint#1#2{\mn@eprint@#1:#2::\@nil}
\def\mn@eprint@arXiv#1{\href {http://arxiv.org/abs/#1} {{\tt arXiv:#1}}}
\def\mn@eprint@dblp#1{\href {http://dblp.uni-trier.de/rec/bibtex/#1.xml}
  {dblp:#1}}
\def\mn@eprint@#1:#2:#3:#4\@nil{\def\@tempa {#1}\def\@tempb {#2}\def\@tempc
  {#3}\ifx \@tempc \@empty \let \@tempc \@tempb \let \@tempb \@tempa \fi \ifx
  \@tempb \@empty \def\@tempb {arXiv}\fi \@ifundefined
  {mn@eprint@\@tempb}{\@tempb:\@tempc}{\expandafter \expandafter \csname
  mn@eprint@\@tempb\endcsname \expandafter{\@tempc}}}

\bibitem[\protect\citeauthoryear{{Cash}}{{Cash}}{1979}]{cas79}
{Cash} W.,  1979, \mn@doi [\apj] {10.1086/156922}, \href
  {http://adsabs.harvard.edu/abs/1979ApJ...228..939C} {228, 939}

\bibitem[\protect\citeauthoryear{{Churazov}, {Forman}, {Jones}  \&
  {B{\"o}hringer}}{{Churazov} et~al.}{2003}]{chu03}
{Churazov} E.,  {Forman} W.,  {Jones} C.,   {B{\"o}hringer} H.,  2003, \mn@doi
  [\apj] {10.1086/374923}, \href
  {https://ui.adsabs.harvard.edu/abs/2003ApJ...590..225C} {590, 225}

\bibitem[\protect\citeauthoryear{{De Grandi} \& {Molendi}}{{De Grandi} \&
  {Molendi}}{2001}]{deg01}
{De Grandi} S.,  {Molendi} S.,  2001, \mn@doi [\apj] {10.1086/320098}, \href
  {https://ui.adsabs.harvard.edu/abs/2001ApJ...551..153D} {551, 153}

\bibitem[\protect\citeauthoryear{{Erdim}, {Ezer}, {{\"U}nver}, {Hazar}  \&
  {Hudaverdi}}{{Erdim} et~al.}{2021}]{erd21}
{Erdim} M.~K.,  {Ezer} C.,  {{\"U}nver} O.,  {Hazar} F.,   {Hudaverdi} M.,
  2021, \mn@doi [\mnras] {10.1093/mnras/stab2730}, \href
  {https://ui.adsabs.harvard.edu/abs/2021MNRAS.508.3337E} {508, 3337}

\bibitem[\protect\citeauthoryear{{Fink} et~al.,}{{Fink} et~al.}{2014}]{fin14}
{Fink} M.,  et~al., 2014, \mn@doi [\mnras] {10.1093/mnras/stt2315}, \href
  {https://ui.adsabs.harvard.edu/abs/2014MNRAS.438.1762F} {438, 1762}

\bibitem[\protect\citeauthoryear{{Fukushima}, {Kobayashi}  \&
  {Matsushita}}{{Fukushima} et~al.}{2022}]{fuk22}
{Fukushima} K.,  {Kobayashi} S.~B.,   {Matsushita} K.,  2022, \mn@doi [\mnras]
  {10.1093/mnras/stac1590}, \href
  {https://ui.adsabs.harvard.edu/abs/2022MNRAS.514.4222F} {514, 4222}

\bibitem[\protect\citeauthoryear{{Gamezo}, {Khokhlov}  \& {Oran}}{{Gamezo}
  et~al.}{2005}]{gam05}
{Gamezo} V.~N.,  {Khokhlov} A.~M.,   {Oran} E.~S.,  2005, \mn@doi [\apj]
  {10.1086/428767}, \href
  {https://ui.adsabs.harvard.edu/abs/2005ApJ...623..337G} {623, 337}

\bibitem[\protect\citeauthoryear{{Gastaldello} et~al.,}{{Gastaldello}
  et~al.}{2010}]{gas10}
{Gastaldello} F.,  et~al., 2010, \mn@doi [\aap] {10.1051/0004-6361/201014279},
  \href {https://ui.adsabs.harvard.edu/abs/2010A&A...522A..34G} {522, A34}

\bibitem[\protect\citeauthoryear{{Gatuzz}, {Sanders}, {Dennerl}, {Pinto},
  {Fabian}, {Tamura}, {Walker}  \& {ZuHone}}{{Gatuzz} et~al.}{2022a}]{gat22a}
{Gatuzz} E.,  {Sanders} J.~S.,  {Dennerl} K.,  {Pinto} C.,  {Fabian} A.~C.,
  {Tamura} T.,  {Walker} S.~A.,   {ZuHone} J.,  2022a, \mn@doi [\mnras]
  {10.1093/mnras/stab2661}, \href
  {https://ui.adsabs.harvard.edu/abs/2022MNRAS.511.4511G} {511, 4511}

\bibitem[\protect\citeauthoryear{{Gatuzz} et~al.,}{{Gatuzz}
  et~al.}{2022b}]{gat22b}
{Gatuzz} E.,  et~al., 2022b, \mn@doi [\mnras] {10.1093/mnras/stac846}, \href
  {https://ui.adsabs.harvard.edu/abs/2022MNRAS.513.1932G} {513, 1932}

\bibitem[\protect\citeauthoryear{{Gatuzz} et~al.,}{{Gatuzz}
  et~al.}{2023a}]{gat23a}
{Gatuzz} E.,  et~al., 2023a, \mn@doi [\mnras] {10.1093/mnras/stad447}, \href
  {https://ui.adsabs.harvard.edu/abs/2023MNRAS.tmp..470G} {}

\bibitem[\protect\citeauthoryear{{Gatuzz}, {Mohapatra}, {Federrath}, {Sanders},
  {Liu}, {Walker}  \& {Pinto}}{{Gatuzz} et~al.}{2023b}]{gat23b}
{Gatuzz} E.,  {Mohapatra} R.,  {Federrath} C.,  {Sanders} J.~S.,  {Liu} A.,
  {Walker} S.~A.,   {Pinto} C.,  2023b, arXiv e-prints, \href
  {https://ui.adsabs.harvard.edu/abs/2023arXiv230702576G} {p. arXiv:2307.02576}

\bibitem[\protect\citeauthoryear{{Gatuzz} et~al.,}{{Gatuzz}
  et~al.}{2023c}]{gat23c}
{Gatuzz} E.,  et~al., 2023c, \mn@doi [\mnras] {10.1093/mnras/stad1132}, \href
  {https://ui.adsabs.harvard.edu/abs/2023MNRAS.522.2325G} {522, 2325}

\bibitem[\protect\citeauthoryear{{Gendron-Marsolais}
  et~al.,}{{Gendron-Marsolais} et~al.}{2017}]{gen17}
{Gendron-Marsolais} M.,  et~al., 2017, \mn@doi [\apj]
  {10.3847/1538-4357/aa8a6f}, \href
  {https://ui.adsabs.harvard.edu/abs/2017ApJ...848...26G} {848, 26}

\bibitem[\protect\citeauthoryear{{Hitomi Collaboration} et~al.,}{{Hitomi
  Collaboration} et~al.}{2018}]{hit18}
{Hitomi Collaboration} et~al., 2018, \mn@doi [\pasj] {10.1093/pasj/psx156},
  \href {https://ui.adsabs.harvard.edu/abs/2018PASJ...70...12H} {70, 12}

\bibitem[\protect\citeauthoryear{{Hoeflich} \& {Khokhlov}}{{Hoeflich} \&
  {Khokhlov}}{1996}]{hoe96}
{Hoeflich} P.,  {Khokhlov} A.,  1996, \mn@doi [\apj] {10.1086/176748}, \href
  {https://ui.adsabs.harvard.edu/abs/1996ApJ...457..500H} {457, 500}

\bibitem[\protect\citeauthoryear{{Iwasawa}, {Fabian}, {Young}, {Inoue}  \&
  {Matsumoto}}{{Iwasawa} et~al.}{1999}]{iwa99}
{Iwasawa} K.,  {Fabian} A.~C.,  {Young} A.~J.,  {Inoue} H.,   {Matsumoto} C.,
  1999, \mn@doi [\mnras] {10.1046/j.1365-8711.1999.02671.x}, \href
  {http://adsabs.harvard.edu/abs/1999MNRAS.306L..19I} {306, L19}

\bibitem[\protect\citeauthoryear{{Kasen} \& {Plewa}}{{Kasen} \&
  {Plewa}}{2007}]{kas07}
{Kasen} D.,  {Plewa} T.,  2007, \mn@doi [\apj] {10.1086/516834}, \href
  {https://ui.adsabs.harvard.edu/abs/2007ApJ...662..459K} {662, 459}

\bibitem[\protect\citeauthoryear{{Lakhchaura}, {Mernier}  \&
  {Werner}}{{Lakhchaura} et~al.}{2019}]{lak19}
{Lakhchaura} K.,  {Mernier} F.,   {Werner} N.,  2019, \mn@doi [\aap]
  {10.1051/0004-6361/201834755}, \href
  {https://ui.adsabs.harvard.edu/abs/2019A&A...623A..17L} {623, A17}

\bibitem[\protect\citeauthoryear{{Liedahl}, {Osterheld}  \&
  {Goldstein}}{{Liedahl} et~al.}{1995}]{lie95}
{Liedahl} D.~A.,  {Osterheld} A.~L.,   {Goldstein} W.~H.,  1995, \mn@doi
  [\apjl] {10.1086/187729}, \href
  {https://ui.adsabs.harvard.edu/abs/1995ApJ...438L.115L} {438, L115}

\bibitem[\protect\citeauthoryear{{Liu}, {Zhai}  \& {Tozzi}}{{Liu}
  et~al.}{2019}]{liu19}
{Liu} A.,  {Zhai} M.,   {Tozzi} P.,  2019, \mn@doi [\mnras]
  {10.1093/mnras/stz533}, \href
  {https://ui.adsabs.harvard.edu/abs/2019MNRAS.485.1651L} {485, 1651}

\bibitem[\protect\citeauthoryear{{Liu}, {Tozzi}, {Ettori}, {De Grandi},
  {Gastaldello}, {Rosati}  \& {Norman}}{{Liu} et~al.}{2020}]{liu20}
{Liu} A.,  {Tozzi} P.,  {Ettori} S.,  {De Grandi} S.,  {Gastaldello} F.,
  {Rosati} P.,   {Norman} C.,  2020, \mn@doi [\aap]
  {10.1051/0004-6361/202037506}, \href
  {https://ui.adsabs.harvard.edu/abs/2020A&A...637A..58L} {637, A58}

\bibitem[\protect\citeauthoryear{{Lodders} \& {Palme}}{{Lodders} \&
  {Palme}}{2009}]{lod09}
{Lodders} K.,  {Palme} H.,  2009, Meteoritics and Planetary Science Supplement,
  \href {https://ui.adsabs.harvard.edu/abs/2009M&PSA..72.5154L} {72, 5154}

\bibitem[\protect\citeauthoryear{{Long} et~al.,}{{Long} et~al.}{2014}]{lon14}
{Long} M.,  et~al., 2014, \mn@doi [\apj] {10.1088/0004-637X/789/2/103}, \href
  {https://ui.adsabs.harvard.edu/abs/2014ApJ...789..103L} {789, 103}

\bibitem[\protect\citeauthoryear{{Lucey}, {Currie}  \& {Dickens}}{{Lucey}
  et~al.}{1986}]{luc86a}
{Lucey} J.~R.,  {Currie} M.~J.,   {Dickens} R.~J.,  1986, \mn@doi [\mnras]
  {10.1093/mnras/221.2.453}, \href
  {https://ui.adsabs.harvard.edu/abs/1986MNRAS.221..453L} {221, 453}

\bibitem[\protect\citeauthoryear{{Madsen}, {Beardmore}, {Forster}, {Guainazzi},
  {Marshall}, {Miller}, {Page}  \& {Stuhlinger}}{{Madsen} et~al.}{2017}]{mad17}
{Madsen} K.~K.,  {Beardmore} A.~P.,  {Forster} K.,  {Guainazzi} M.,  {Marshall}
  H.~L.,  {Miller} E.~D.,  {Page} K.~L.,   {Stuhlinger} M.,  2017, \mn@doi
  [\aj] {10.3847/1538-3881/153/1/2}, \href
  {http://adsabs.harvard.edu/abs/2017AJ....153....2M} {153, 2}

\bibitem[\protect\citeauthoryear{{Mantz}, {Allen}, {Morris}, {Simionescu},
  {Urban}, {Werner}  \& {Zhuravleva}}{{Mantz} et~al.}{2017}]{man17}
{Mantz} A.~B.,  {Allen} S.~W.,  {Morris} R.~G.,  {Simionescu} A.,  {Urban} O.,
  {Werner} N.,   {Zhuravleva} I.,  2017, \mn@doi [\mnras]
  {10.1093/mnras/stx2200}, \href
  {https://ui.adsabs.harvard.edu/abs/2017MNRAS.472.2877M} {472, 2877}

\bibitem[\protect\citeauthoryear{{Markevitch} et~al.,}{{Markevitch}
  et~al.}{2000}]{mar00}
{Markevitch} M.,  et~al., 2000, \mn@doi [\apj] {10.1086/309470}, \href
  {https://ui.adsabs.harvard.edu/abs/2000ApJ...541..542M} {541, 542}

\bibitem[\protect\citeauthoryear{{Markowitz} et~al.,}{{Markowitz}
  et~al.}{2007}]{mar07}
{Markowitz} A.,  et~al., 2007, \mn@doi [\apj] {10.1086/519271}, \href
  {http://adsabs.harvard.edu/abs/2007ApJ...665..209M} {665, 209}

\bibitem[\protect\citeauthoryear{{Matsushita}}{{Matsushita}}{2011}]{mat11}
{Matsushita} K.,  2011, \mn@doi [\aap] {10.1051/0004-6361/200913432}, \href
  {https://ui.adsabs.harvard.edu/abs/2011A&A...527A.134M} {527, A134}

\bibitem[\protect\citeauthoryear{{Matsushita}, {B{\"o}hringer}, {Takahashi}  \&
  {Ikebe}}{{Matsushita} et~al.}{2007}]{mat07}
{Matsushita} K.,  {B{\"o}hringer} H.,  {Takahashi} I.,   {Ikebe} Y.,  2007,
  \mn@doi [\aap] {10.1051/0004-6361:20041577}, \href
  {https://ui.adsabs.harvard.edu/abs/2007A&A...462..953M} {462, 953}

\bibitem[\protect\citeauthoryear{{Mernier} et~al.,}{{Mernier}
  et~al.}{2016}]{mer16}
{Mernier} F.,  et~al., 2016, \mn@doi [\aap] {10.1051/0004-6361/201628765},
  \href {https://ui.adsabs.harvard.edu/abs/2016A&A...595A.126M} {595, A126}

\bibitem[\protect\citeauthoryear{{Mernier} et~al.,}{{Mernier}
  et~al.}{2017}]{mer17}
{Mernier} F.,  et~al., 2017, \mn@doi [\aap] {10.1051/0004-6361/201630075},
  \href {https://ui.adsabs.harvard.edu/abs/2017A&A...603A..80M} {603, A80}

\bibitem[\protect\citeauthoryear{{Mernier} et~al.,}{{Mernier}
  et~al.}{2018}]{mer18}
{Mernier} F.,  et~al., 2018, \mn@doi [\ssr] {10.1007/s11214-018-0565-7}, \href
  {https://ui.adsabs.harvard.edu/abs/2018SSRv..214..129M} {214, 129}

\bibitem[\protect\citeauthoryear{{Mernier} et~al.,}{{Mernier}
  et~al.}{2020}]{mer20}
{Mernier} F.,  et~al., 2020, \mn@doi [\aap] {10.1051/0004-6361/202038638},
  \href {https://ui.adsabs.harvard.edu/abs/2020A&A...642A..90M} {642, A90}

\bibitem[\protect\citeauthoryear{{Mulchaey}, {Kasliwal}  \&
  {Kollmeier}}{{Mulchaey} et~al.}{2014}]{mul14}
{Mulchaey} J.~S.,  {Kasliwal} M.~M.,   {Kollmeier} J.~A.,  2014, \mn@doi
  [\apjl] {10.1088/2041-8205/780/2/L34}, \href
  {https://ui.adsabs.harvard.edu/abs/2014ApJ...780L..34M} {780, L34}

\bibitem[\protect\citeauthoryear{{Nomoto}, {Kobayashi}  \& {Tominaga}}{{Nomoto}
  et~al.}{2013}]{nom13}
{Nomoto} K.,  {Kobayashi} C.,   {Tominaga} N.,  2013, \mn@doi [\araa]
  {10.1146/annurev-astro-082812-140956}, \href
  {https://ui.adsabs.harvard.edu/abs/2013ARA&A..51..457N} {51, 457}

\bibitem[\protect\citeauthoryear{{Panagoulia}, {Fabian}  \&
  {Sanders}}{{Panagoulia} et~al.}{2013}]{pan13}
{Panagoulia} E.~K.,  {Fabian} A.~C.,   {Sanders} J.~S.,  2013, \mn@doi [\mnras]
  {10.1093/mnras/stt969}, \href
  {https://ui.adsabs.harvard.edu/abs/2013MNRAS.433.3290P} {433, 3290}

\bibitem[\protect\citeauthoryear{{Panagoulia}, {Sanders}  \&
  {Fabian}}{{Panagoulia} et~al.}{2015}]{pan15}
{Panagoulia} E.~K.,  {Sanders} J.~S.,   {Fabian} A.~C.,  2015, \mn@doi [\mnras]
  {10.1093/mnras/stu2469}, \href
  {https://ui.adsabs.harvard.edu/abs/2015MNRAS.447..417P} {447, 417}

\bibitem[\protect\citeauthoryear{{Plewa}}{{Plewa}}{2007}]{ple07}
{Plewa} T.,  2007, \mn@doi [\apj] {10.1086/511412}, \href
  {https://ui.adsabs.harvard.edu/abs/2007ApJ...657..942P} {657, 942}

\bibitem[\protect\citeauthoryear{{Plewa}, {Calder}  \& {Lamb}}{{Plewa}
  et~al.}{2004}]{ple04}
{Plewa} T.,  {Calder} A.~C.,   {Lamb} D.~Q.,  2004, \mn@doi [\apjl]
  {10.1086/424036}, \href
  {https://ui.adsabs.harvard.edu/abs/2004ApJ...612L..37P} {612, L37}

\bibitem[\protect\citeauthoryear{{Roediger}, {Kraft}, {Nulsen}, {Churazov},
  {Forman}, {Br{\"u}ggen}  \& {Kokotanekova}}{{Roediger} et~al.}{2013}]{roed13}
{Roediger} E.,  {Kraft} R.~P.,  {Nulsen} P.,  {Churazov} E.,  {Forman} W.,
  {Br{\"u}ggen} M.,   {Kokotanekova} R.,  2013, \mn@doi [\mnras]
  {10.1093/mnras/stt1691}, \href
  {https://ui.adsabs.harvard.edu/abs/2013MNRAS.436.1721R} {436, 1721}

\bibitem[\protect\citeauthoryear{{R{\"o}pke} et~al.,}{{R{\"o}pke}
  et~al.}{2012}]{rop12}
{R{\"o}pke} F.~K.,  et~al., 2012, \mn@doi [\apjl]
  {10.1088/2041-8205/750/1/L19}, \href
  {https://ui.adsabs.harvard.edu/abs/2012ApJ...750L..19R} {750, L19}

\bibitem[\protect\citeauthoryear{{Sakuma}, {Ota}, {Sato}, {Sato}  \&
  {Matsushita}}{{Sakuma} et~al.}{2011}]{sak11}
{Sakuma} E.,  {Ota} N.,  {Sato} K.,  {Sato} T.,   {Matsushita} K.,  2011,
  \mn@doi [\pasj] {10.1093/pasj/63.sp3.S979}, \href
  {https://ui.adsabs.harvard.edu/abs/2011PASJ...63S.979S} {63, S979}

\bibitem[\protect\citeauthoryear{{Sanders} \& {Fabian}}{{Sanders} \&
  {Fabian}}{2002}]{san02}
{Sanders} J.~S.,  {Fabian} A.~C.,  2002, \mn@doi [\mnras]
  {10.1046/j.1365-8711.2002.05211.x}, \href
  {https://ui.adsabs.harvard.edu/abs/2002MNRAS.331..273S} {331, 273}

\bibitem[\protect\citeauthoryear{{Sanders} \& {Fabian}}{{Sanders} \&
  {Fabian}}{2006a}]{san06c}
{Sanders} J.~S.,  {Fabian} A.~C.,  2006a, \mn@doi [\mnras]
  {10.1111/j.1365-2966.2006.10497.x}, \href
  {https://ui.adsabs.harvard.edu/abs/2006MNRAS.370...63S} {370, 63}

\bibitem[\protect\citeauthoryear{{Sanders} \& {Fabian}}{{Sanders} \&
  {Fabian}}{2006b}]{san06b}
{Sanders} J.~S.,  {Fabian} A.~C.,  2006b, \mn@doi [\mnras]
  {10.1111/j.1365-2966.2006.10779.x}, \href
  {https://ui.adsabs.harvard.edu/abs/2006MNRAS.371.1483S} {371, 1483}

\bibitem[\protect\citeauthoryear{{Sanders} et~al.,}{{Sanders}
  et~al.}{2016}]{san16}
{Sanders} J.~S.,  et~al., 2016, \mn@doi [\mnras] {10.1093/mnras/stv2972}, \href
  {https://ui.adsabs.harvard.edu/abs/2016MNRAS.457...82S} {457, 82}

\bibitem[\protect\citeauthoryear{{Sanders} et~al.,}{{Sanders}
  et~al.}{2020}]{san20}
{Sanders} J.~S.,  et~al., 2020, \mn@doi [\aap] {10.1051/0004-6361/201936468},
  \href {https://ui.adsabs.harvard.edu/abs/2020A&A...633A..42S} {633, A42}

\bibitem[\protect\citeauthoryear{{Schellenberger}, {Reiprich}, {Lovisari},
  {Nevalainen}  \& {David}}{{Schellenberger} et~al.}{2015}]{sch15}
{Schellenberger} G.,  {Reiprich} T.~H.,  {Lovisari} L.,  {Nevalainen} J.,
  {David} L.,  2015, \mn@doi [\aap] {10.1051/0004-6361/201424085}, \href
  {https://ui.adsabs.harvard.edu/abs/2015A&A...575A..30S} {575, A30}

\bibitem[\protect\citeauthoryear{{Seitenzahl} et~al.,}{{Seitenzahl}
  et~al.}{2013}]{sei13}
{Seitenzahl} I.~R.,  et~al., 2013, \mn@doi [\mnras] {10.1093/mnras/sts402},
  \href {https://ui.adsabs.harvard.edu/abs/2013MNRAS.429.1156S} {429, 1156}

\bibitem[\protect\citeauthoryear{{Simionescu}, {Werner}, {Urban}, {Allen},
  {Ichinohe}  \& {Zhuravleva}}{{Simionescu} et~al.}{2015}]{sim15}
{Simionescu} A.,  {Werner} N.,  {Urban} O.,  {Allen} S.~W.,  {Ichinohe} Y.,
  {Zhuravleva} I.,  2015, \mn@doi [\apjl] {10.1088/2041-8205/811/2/L25}, \href
  {https://ui.adsabs.harvard.edu/abs/2015ApJ...811L..25S} {811, L25}

\bibitem[\protect\citeauthoryear{{Simionescu} et~al.,}{{Simionescu}
  et~al.}{2019}]{sim19}
{Simionescu} A.,  et~al., 2019, \mn@doi [\mnras] {10.1093/mnras/sty3220}, \href
  {https://ui.adsabs.harvard.edu/abs/2019MNRAS.483.1701S} {483, 1701}

\bibitem[\protect\citeauthoryear{{Str{\"u}der} et~al.,}{{Str{\"u}der}
  et~al.}{2001}]{str01}
{Str{\"u}der} L.,  et~al., 2001, \mn@doi [\aap] {10.1051/0004-6361:20000066},
  \href {http://adsabs.harvard.edu/abs/2001A%26A...365L..18S} {365, L18}

\bibitem[\protect\citeauthoryear{{Takahashi} et~al.,}{{Takahashi}
  et~al.}{2009}]{tak09}
{Takahashi} I.,  et~al., 2009, \mn@doi [\apj] {10.1088/0004-637X/701/1/377},
  \href {https://ui.adsabs.harvard.edu/abs/2009ApJ...701..377T} {701, 377}

\bibitem[\protect\citeauthoryear{{Urban}, {Werner}, {Allen}, {Simionescu}  \&
  {Mantz}}{{Urban} et~al.}{2017}]{urb17}
{Urban} O.,  {Werner} N.,  {Allen} S.~W.,  {Simionescu} A.,   {Mantz} A.,
  2017, \mn@doi [\mnras] {10.1093/mnras/stx1542}, \href
  {https://ui.adsabs.harvard.edu/abs/2017MNRAS.470.4583U} {470, 4583}

\bibitem[\protect\citeauthoryear{{Vijayan} \& {Li}}{{Vijayan} \&
  {Li}}{2022}]{vij22}
{Vijayan} A.,  {Li} M.,  2022, \mn@doi [\mnras] {10.1093/mnras/stab3413}, \href
  {https://ui.adsabs.harvard.edu/abs/2022MNRAS.510..568V} {510, 568}

\bibitem[\protect\citeauthoryear{{Vikhlinin}, {Markevitch}  \&
  {Murray}}{{Vikhlinin} et~al.}{2001}]{vik01}
{Vikhlinin} A.,  {Markevitch} M.,   {Murray} S.~S.,  2001, \mn@doi [\apj]
  {10.1086/320078}, \href
  {https://ui.adsabs.harvard.edu/abs/2001ApJ...551..160V} {551, 160}

\bibitem[\protect\citeauthoryear{{Walker}, {Fabian}  \& {Sanders}}{{Walker}
  et~al.}{2013}]{wal13a}
{Walker} S.~A.,  {Fabian} A.~C.,   {Sanders} J.~S.,  2013, \mn@doi [\mnras]
  {10.1093/mnras/stt1515}, \href
  {https://ui.adsabs.harvard.edu/abs/2013MNRAS.435.3221W} {435, 3221}

\bibitem[\protect\citeauthoryear{{Werner}, {Durret}, {Ohashi}, {Schindler}  \&
  {Wiersma}}{{Werner} et~al.}{2008}]{wer08}
{Werner} N.,  {Durret} F.,  {Ohashi} T.,  {Schindler} S.,   {Wiersma} R.~P.~C.,
   2008, \mn@doi [\ssr] {10.1007/s11214-008-9320-9}, \href
  {https://ui.adsabs.harvard.edu/abs/2008SSRv..134..337W} {134, 337}

\bibitem[\protect\citeauthoryear{{Werner}, {Urban}, {Simionescu}  \&
  {Allen}}{{Werner} et~al.}{2013}]{wer13}
{Werner} N.,  {Urban} O.,  {Simionescu} A.,   {Allen} S.~W.,  2013, \mn@doi
  [\nat] {10.1038/nature12646}, \href
  {https://ui.adsabs.harvard.edu/abs/2013Natur.502..656W} {502, 656}

\bibitem[\protect\citeauthoryear{{Wilms}, {Allen}  \& {McCray}}{{Wilms}
  et~al.}{2000}]{wil00}
{Wilms} J.,  {Allen} A.,   {McCray} R.,  2000, \mn@doi [\apj] {10.1086/317016},
  \href {http://adsabs.harvard.edu/abs/2000ApJ...542..914W} {542, 914}

\bibitem[\protect\citeauthoryear{{Yoshino} et~al.,}{{Yoshino}
  et~al.}{2009}]{yos09}
{Yoshino} T.,  et~al., 2009, \mn@doi [\pasj] {10.1093/pasj/61.4.805}, \href
  {https://ui.adsabs.harvard.edu/abs/2009PASJ...61..805Y} {61, 805}

\bibitem[\protect\citeauthoryear{{de Plaa}, {Werner}, {Bleeker}, {Vink},
  {Kaastra}  \& {M{\'e}ndez}}{{de Plaa} et~al.}{2007}]{dep07}
{de Plaa} J.,  {Werner} N.,  {Bleeker} J.~A.~M.,  {Vink} J.,  {Kaastra} J.~S.,
   {M{\'e}ndez} M.,  2007, \mn@doi [\aap] {10.1051/0004-6361:20066382}, \href
  {https://ui.adsabs.harvard.edu/abs/2007A&A...465..345D} {465, 345}

\makeatother
\end{thebibliography}
 \newcommand{\noop}[1]{}

\end{document}